\documentclass[manuscript,screen,nonacm]{acmart}

\setcopyright{none}

\renewcommand\footnotetextcopyrightpermission[1]{}
\AtBeginDocument{%
  }

\usepackage{tabularx}
\usepackage{booktabs}
\usepackage{array}

\usepackage{subcaption}
\usepackage{multirow}
\usepackage{graphicx}

\usepackage[flushleft]{threeparttable}  

\allowdisplaybreaks[1] 

\newcommand{\descr}[1]{\vspace{0.2cm} \noindent \textbf{#1}}

\setcopyright{none}
\acmConference{}
\acmBooktitle{}
\acmYear{}
\acmDOI{}
\acmISBN{}

\begin{document}

\title{Johnny Still Receives Spam SMS: Assessing the Robustness of SMS Spam Detection}

\author{Muhammad Salman}
\affiliation{%
  \institution{Macquarie University}
  \city{Sydney}
  \country{Australia}
}
\email{muhammad.salman2@students.mq.edu.au}

\author{Muhammad Islam}
\affiliation{%
  \institution{University of Engineering and Technology Mardan}
  \city{Mardan}
  \country{Pakistan}
}
\email{imisl4m@gmail.com}

\author{Muhammad Ikram}
\affiliation{%
  \institution{Macquarie University}
  \city{Sydney}
  \country{Australia}
}
\email{muhammad.ikram@mq.edu.au}

\author{Mohamed Ali Kaafar}
\affiliation{%
  \institution{Macquarie University}
  \city{Sydney}
  \country{Australia}
}
\email{dali.kaafar@mq.edu.au}

\renewcommand{\shortauthors}{Salman et al.}

\begin{abstract}
SMS spam detection systems often achieve high accuracy in controlled environments but struggle against adversarial attacks and increasingly sophisticated spam tactics in real-world deployments. In this paper, we evaluate the robustness of SMS anti-spam systems that end users actually rely on, including commercial messaging applications, third-party anti-spam services, and publicly available open-weight models hosted on Hugging Face. We evaluate these systems under both standard and adversarial conditions, considering perceptible and state-of-the-art imperceptible attacks. We include only perturbations that we verify survive real SMS or RCS delivery, rather than lab-only artifacts.

Our experiments reveal significant gaps in existing spam detectors' ability to identify adversarially manipulated messages. We further demonstrate that adversarial training alone is insufficient. Using an explicit held-out evaluation protocol, we find that robustness transfers well within a perturbation family but degrades sharply against structurally distinct, encoding-level attacks. To address these weaknesses, we propose a multi-model ensemble that combines adversarial training with spam classifiers diverse in architecture and tokenization. Our results show that this ensemble, particularly when using a minority-voting strategy, substantially improves robustness against both perceptible and imperceptible adversarial attacks while maintaining competitive classification accuracy. We also characterize the resulting precision--recall trade-off and recommend operating points for false-positive-sensitive and recall-critical deployments. These findings highlight the need for comprehensive robustness evaluations and ensemble-based defenses for building more secure SMS spam detection systems in real-world settings.
\end{abstract}



\maketitle

\section{Introduction}
Despite significant advancements in SMS spam\footnote{In this work, SMS spam refers to any unsolicited message, including advertisements, proselytizing, and scams. We treat phishing/smishing as a subset of spam (a single positive class) rather than a separate category; any message meeting this unsolicited-content criterion is labeled spam.} detection, where state-of-the-art classifiers spanning both conventional machine learning and deep learning (DL) approaches exhibit impressive true-positive rates (TPRs) exceeding 99\% and false-positive rates (FPRs) below 1\% \cite{sahmoud2022spam,sheikhi2020effective,krishna2022spam,xia2020discrete,delany2012sms,roy2020deep}, the real-world effectiveness of these models remains questionable. In practice, these models are frequently deployed within various anti-spam infrastructures, including mobile text applications and commercial anti-spam services. However, they often fail to deliver the expected performance, particularly when confronted with sophisticated spam techniques, evolving spam characteristics—referred to as concept drift and adversarial attacks that exploit the classifiers' vulnerabilities. This discrepancy between controlled experimental success and practical deployment failure highlights the critical need for a deeper investigation into the causes of these failures and the development of more robust solutions.

While periodic retraining of machine learning models is a common practice to address concept drift \cite{costa2014concept,lu2018learning,muller2020addressing}, the issue of robustness remains a critical challenge in SMS spam. Despite its importance, robustness in these classifiers has been insufficiently explored in the existing literature, presenting a significant risk given the adversarial nature of the environments in which these models must operate. Although previous research has partially examined the robustness of spam classifiers against adversarial examples (AEs) \cite{salman2024investigating,li2024spamdam} and the attacks that generate them, known as adversarial attacks, this exploration has been far from comprehensive. Most studies on black-box attacks in SMS spam detection have focused on character perturbations to create AEs, demonstrating their impact on non-hardened spam classifiers and subsequently measuring the improved robustness of models following adversarial training (AT) \cite{li2024spamdam}. This approach typically involves retraining classifiers with datasets augmented by spam messages generated through specific attacks. However, such methods often overlook the broader issue of general robustness, especially against unknown attacks, which is crucial for practical and secure deployments.

The current robustness evaluations in research often fail to replicate the complexity of real-world conditions. Typically, studies either introduce new attacks to showcase classifier vulnerabilities or rely on a single attack for adversarial training, subsequently evaluating robustness against that same attack. In some cases, robustness is assessed against different attacks, but this is usually limited to just one or two additional attacks of the same category (without considering the inherent similarities or differences between the attacks) following the initial black-box attack for AT. This narrow focus is inadequate for assessing general robustness. In practical settings, a classifier should be hardened against a wide spectrum of known attacks, as improving robustness against one attack does not guarantee defense against others. Moreover, unknown attacks—those not encountered during the model’s training or hardening process—pose an even greater challenge. Therefore, a more comprehensive evaluation of robustness, ideally against a diverse array of both known and unknown attacks, is necessary to ensure the reliability and security of SMS spam detection systems in real-world applications.

Furthermore, while it is well-established that machine learning models are vulnerable to adversarial examples, recent studies have revealed that the transferability of these examples in text is asymmetric~\cite{yuan2020transferability}. Specifically, an adversarial example designed to deceive one model (Model A) may not necessarily have the same effect on another model (Model B), and vice versa. This asymmetry is influenced by various factors, including network architecture, tokenization schemes, word embeddings, and model capacity. Among these, tokenization schemes have been found to have the greatest impact on transferability, followed by network architecture, embedding type, and model capacity~\cite{yuan2020transferability}. These findings are particularly encouraging as they suggest that an ensemble approach—employing multiple models with diverse architectures and tokenization strategies—could enhance the overall robustness of spam detection systems. By reducing the likelihood that a single adversarial example will succeed across multiple models, this ensemble approach, combined with adversarial training, could provide a more resilient defense against a wide range of attacks.

To mitigate these vulnerabilities, our research explores potential solutions, including the use of ensemble methods and adversarial training. By leveraging an ensemble of diverse models or incorporating adversarial training into the development process, we aim to enhance the robustness of these systems, making them more resilient to both known and unknown adversarial strategies.

In our work, we address these gaps with the following contributions:

\begin{itemize}
    \item \textbf{Analysis of Anti-spam Infrastructure and Open-Weight Spam Models.} We conducted an extensive series of experiments evaluating the anti-spam infrastructure end-users actually rely on---text messaging apps and third-party anti-spam services---together with publicly available pen-weight spam models hosted on HuggingFace, under both benign and adversarial conditions. To do so, we transmitted a substantial number of benign and spam messages from a freshly registered SIM through APIs to third-party services and to messaging apps on smartphones under our control, and separately implemented and tested the open-weight models. None of the evaluated systems identified all spam messages: false-negative rates (FNR) exceeded 5\%---our comparative benchmark across systems (Section~\ref{sec:imbalance})---highlighting the inadequacy of current SMS spam detection. All tested systems further proved vulnerable to adversarial manipulation, and we verified that our imperceptible perturbations survive real SMS/RCS delivery (Section~\ref{sec:deliverability}) rather than being lab-only artifacts. We also manually labeled the misdetected messages into 12 spam categories to characterize which spam types evade detection.
   
    \item \textbf{Training, Hardening, and Quantifying Generalization to
    Held-Out Attacks.} To improve SMS spam detection against adversarial
    attacks, we trained our own ML models on a recent SMS spam dataset and
    hardened them through adversarial training, which substantially improved
    robustness against the attacks seen during training. We then defined an
    explicit held-in/held-out protocol to measure whether this robustness
    generalizes, and quantified the generalization gap per attack family. We
    find that robustness transfers almost completely to same-class
    perturbations but leaves a substantial gap against mechanistically
    distinct, encoding-level attacks such as homoglyph substitution---showing
    that adversarial training's generalization is class-bounded rather than
    uniform.

    \item \textbf{Ensemble of Detectors to Mitigate Adversarial Attacks.} We then
    investigate whether an ensemble of detectors can effectively mitigate the
    impact of adversarial attacks by employing different decision rules:
    minority, majority, and consensus. We evaluate various ensembles comprising
    both open-weight models and our own trained models, analyzing their
    performance when classifying both adversarially transformed spam and original
    spam messages. Our findings reveal that an ensemble of our own models,
    particularly when incorporating an adversarially trained model and utilizing a
    minority voting strategy, is highly effective at countering adversarial
    attacks, demonstrating robustness against known adversarial tactics as well as
    a considerable degree of resilience to unknown attacks. Finally, we translate
    these results into deployment guidance: ensemble size and decision rule
    control a recall--precision trade-off, and we recommend operating points for
    false-positive-sensitive traffic (OTP/banking) versus recall-critical traffic
    (marketing-heavy), acknowledging that the highest-recall configurations come
    at the cost of elevated false positives.
\end{itemize}

\section{Background and Related Work}
\subsection{SMS Message Flow and Filtering Points}
To situate our evaluation within the operational SMS ecosystem, we first
outline the path a message traverses from sender to recipient and the points
at which spam filtering can occur. A message originates at the sending
(A-party) device as a Mobile-Originated (MO) submission to the sender's
\textit{Short Message Service Center} (SMSC)~\cite{yatebts_smsflow}. Because
SMS is a store-and-forward service, the SMSC queries the recipient's
\textit{Home Location Register} (HLR) for routing information (SRI-SM) before
performing the Mobile-Terminated (MT) delivery toward the recipient
operator~\cite{yatebts_smsflow, endtoend_smsflow}. The SMSC also performs
protocol conversion between application-facing submission protocols (SMPP) and
inter-operator signaling (SS7/MAP)~\cite{endtoend_smsflow}. At the recipient
operator, a carrier-grade \textit{SMS firewall} typically inspects the
traffic---often via \textit{SMS home routing}, which forces MT messages
through the operator's own platform for inspection---before the message
reaches the recipient (B-party) device and any on-device filtering
application~\cite{tecways_smsfirewall, comcode_smsfirewall}.

Filtering can be applied at three broad layers:
\begin{itemize}
  \item \textbf{Network/operator level.} SMS firewalls inspect signaling and
  routing metadata (e.g., SCCP/MAP consistency, grey-route detection,
  sender-ID spoofing) and enforce volumetric and similarity-based rules on Application-to-Person (A2P) and Person-to-Person (P2P) traffic, largely following GSMA recommendations~\cite{tecways_smsfirewall, comcode_smsfirewall}. In practice, the bulk of commercial spam and grey-route abuse is A2P traffic, so operator firewalls concentrate their controls there~\cite{comcode_smsfirewall}.
  \item \textbf{Gateway/content level.} Operators increasingly augment
  signaling checks with content-based filtering and SMS home routing, so that message content is analyzed before delivery~\cite{tieto_smsfirewall}.
  \item \textbf{On-device/application level.} Mobile messaging and anti-spam
  apps classify messages on the recipient device, independent of the carrier.
  These apps employ a mix of techniques: content-based classification
  (on-device ML models operating on message text)~\cite{google_spamprotection},
  sender reputation and community-sourced number blacklists, and URL/link
  reputation lookups~\cite{google_spamprotection}.
\end{itemize}

Our study deliberately targets the \textbf{on-device/application and
content-based service layers}: the mobile apps and third-party APIs we
evaluate perform content-based filtering only, without access to network
signaling, routing metadata, or operator-level controls. This scoping is also
forward-looking: Rich Communication Services (RCS) traffic is IP-based and
bypasses the SS7/SMSC path entirely, relying more heavily on on-device and
server-side content classification of the kind we
study~\cite{google_spamprotection}. By registering a fresh SIM with no prior
sending history, we neutralize sender-reputation and volumetric network-level
controls, ensuring that any classification we observe is attributable to
content-based analysis rather than metadata or blacklist signals. This design
isolates the content-based detection behavior we intend to measure.

\subsection{Prior Work on SMS Spam and Smishing Detection}
Recent research underscores the challenges and vulnerabilities in SMS spam detection systems, particularly in adversarial and real-world scenarios. Salman et al. \cite{salman2024investigating} demonstrate that SMS spam classifiers, including RoBERTa, perform impressively on controlled datasets but exhibit weaknesses against adversarial manipulations.  Their results suggest that even state-of-the-art models can be circumvented by relatively simple evasion tactics, emphasizing the importance of robustness evaluations. Similarly, Liu et al. \cite{liu2021spam} introduce a spam transformer model optimized for SMS spam detection, showcasing high accuracy but limited exploration of adversarial resilience. Tida and Hsu \cite{tida2022universal} extend this line of work, leveraging transfer learning from pre-trained BERT models to achieve universal spam detection but focus predominantly on cross-domain adaptability rather than adversarial robustness. The aforementioned work aligns with the findings of Morris et al. \cite{morris2020textattack}, who introduced the TextAttack framework, revealing that deep learning text classifiers are highly vulnerable to adversarial text attacks that induce misclassification. Similarly, Gao et al. \cite{gao2018black} confirm that adversarial text perturbations can systematically lead to misclassification in black-box scenarios, further highlighting the vulnerabilities inherent in spam detectors.

The existing literature primarily focuses on developing new attack methods, often without thoroughly investigating countermeasures that could significantly improve the robustness of spam detection systems. Nahapetyan et al. \cite{nahapetyan2024sms} analyze the phishing tactics and underlying infrastructure employed in SMS spam campaigns, offering valuable context for understanding the evolving threat landscape. However, their work does not extend to integrating such insights into practical detection mechanisms. Li et al.~\cite{li2024spamdam} evaluated the robustness of an adversarially trained BERT model against state-of-the-art imperceptible attacks. These attacks, introduced by Boucher et al. \cite{boucher2022bad}, generate adversarial examples by either inserting non-printable Unicode characters or replacing existing characters with homoglyphs, rendering the modifications nearly invisible. Li et al.'s study highlighted the vulnerability of non-hardened spam classifiers to such attacks and demonstrated the effectiveness of adversarial training (AT) in improving model robustness by retraining classifiers on datasets augmented with all types of imperceptibly manipulated spam messages under evaluation. Their evaluation, however, is confined to models tested in isolation: it does not assess the deployed anti-spam infrastructure end-users actually rely on, nor does it quantify how robustness transfers to held-out attack families excluded from adversarial training---both of which are crucial for practical and secure deployments.

\subsection{Gap and Positioning}
Taken together, prior work leaves three gaps. First, studies that evaluate deployed SMS anti-spam infrastructure do so only under \textit{benign} conditions: Tang et al.~\cite{tang2022clues} measure anti-spam services, bulk SMS services, and messaging apps against real reported spam and find substantial false positives and missed detections, but do not subject these systems to adversarial perturbation. Second, work on adversarial robustness in SMS evaluates models \textit{in isolation} rather than as deployed systems---Li et al.~\cite{li2024spamdam} rigorously analyze adversarial and poisoning resistance of their own trained detectors, and Salman et al.~\cite{salman2024investigating} show classifier vulnerability to evasion, but neither tests the on-device apps and third-party services end-users actually rely on. Third, adversarial
training is typically applied against known attack families, without assessing whether the resulting robustness transfers to \textit{unseen} tactics. Our work closes these gaps by (a) measuring real-world on-device apps and third-party services alongside open-weight models, (b) evaluating them under both perceptible and imperceptible adversarial conditions rather than benign spam alone, and (c) proposing an ensemble defense whose generalization to held-out attack families we quantify
explicitly.

To clearly position our work relative to existing literature, we contrast our proposed ensemble defense with prior NLP-adversarial ensemble defenses along four key dimensions. First, regarding \textit{evaluation scope}, while existing ensembles (e.g.,~\cite{dedhia2022sms,bilgen2024egma,liu2024stacking}) evaluate only isolated models in local environments, we assess actual deployed anti-spam infrastructure (on-device mobile applications and third-party services) that end-users rely on. Second, concerning the \textit{decision rule}, we evaluate minority, majority, and consensus voting, contrasting with standard defenses that rely exclusively on majority or soft voting, which we show are highly fragile under adversarial conditions. Third, for \textit{member selection}, our ensemble leverages asymmetric transferability in text to select models with diverse architectures and tokenization schemes, rather than simple homogeneous model combinations. Fourth, in terms of \textit{threat coverage}, we evaluate both perceptible and deliverable imperceptible attacks, and explicitly measure performance transfer to held-out, unseen attack families, which is typically omitted in generic-text or SMS-specific ensemble designs.

\section{Materials and Methods}
In the following sections, we outline our research methodologies and evaluation setup. We begin with an overview of the anti-spam infrastructure assessed, followed by a description of the dataset used. We then discuss the evaluation metrics and threat model that frame our experiments. Next, we detail the adversarial attacks implemented to challenge the spam detectors and conclude with the adversarial training techniques employed to enhance model robustness.

\subsection{Anti-Spam Infrastructure and Open-Weight Models}

\descr{Spam filtering Mobile Text apps.}
Several text messaging apps in the Google Play Store and Apple App Store incorporate anti-spam features. We selected three text messaging apps on \textit{Android} (Mezo~\cite{mezo}, Version 17.0.41) and \textit{iOS} (VeroSMS~\cite{verosms}, Version 1.5.3; SMS Shield~\cite{smsshield}, Version 1.1.1) based on their content-based spam filtering, popularity, and previous research~\cite{tang2022clues}. For each app, we used its default settings, which employ machine learning to automatically identify SMS spam; no manual keywords or custom filtering rules were provided. For our experiments, we utilized three cellphones: the iPhone 15 Pro Max (iOS version 26.5) and Samsung Galaxy A52 (Android version 14, One UI 6.1) as recipients, and the Samsung Galaxy S24 Ultra (Android version 14, One UI 6.1) as the spam sender. The Google Messages app was employed as the text messaging platform on the sender's phone. To automate the message-sending process, we used Selenium~\cite{selend}. Given the different graphical user interfaces (GUIs) of the text messaging apps, we manually inspected each app to check if it flagged or blocked incoming spam messages, treating a message as detected if it was marked as spam by the app. The outcome for each message was manually recorded in an Excel sheet. To avoid detection through blacklists and network-level filtering, we registered a new SIM card to send spam SMSes and manually confirmed if each message reached the recipient. To ensure reproducibility, we summarize the app/service versions, OS/device details, exact settings, definition of detection (flagged/blocked), and logging method in Table~\ref{tab:reproducibility-apps}.

\begin{table}[hbt!]
\centering
\caption{Reproducibility details of evaluated mobile anti-spam text applications.}
\label{tab:reproducibility-apps}
\small
\begin{tabular}{p{2.5cm} p{3.2cm} p{2.8cm} p{2.2cm} p{2.2cm}}
\toprule
\textbf{App (Version)} & \textbf{OS \& Device} & \textbf{Settings} & \textbf{Flagged State} & \textbf{Logging Method} \\
\midrule
SMS Shield (1.1.1) & iOS 26.5 on iPhone 15 Pro Max & Default ML auto-filtering & Marked/flagged as spam in UI & Manual Excel logging \\
VeroSMS (1.5.3) & iOS 26.5 on iPhone 15 Pro Max & Default ML auto-filtering & Marked/flagged as spam in UI & Manual Excel logging \\
Mezo (17.0.41) & Android 14 (One UI 6.1) on Samsung Galaxy A52 & Default ML auto-filtering & Marked/flagged as spam in UI & Manual Excel logging \\
\bottomrule
\end{tabular}
\end{table}

\descr{Third party anti-spam services.}
We conducted an extensive search to identify third-party online anti-spam SMS services that utilize content-based filtering and machine learning to detect spam texts. Here, we distinguish \textit{dedicated SMS anti-spam services}---purpose-built for the SMS channel---from \textit{general-purpose text spam services}, which are designed for email or generic text and applied to SMS in our setting. To our knowledge, there are currently no dedicated SMS anti-spam services, but several general-purpose text spam detection services exist, such as Spam Hunter~\cite{spamhunter}, OOPSpam~\cite{oopspam}, and Plino~\cite{plino}. These services provide APIs that accept text input and return detection results. Among them, Plino offers a text classification feature that labels content as either ``spam'' or ``ham''. Conversely, Spam Hunter and OOPSpam provide a spam probability score ranging from ``0 to 10'' and ``0 to 6'', respectively. For classification purposes, we applied the vendor-recommended threshold values of ``greater than 6'' for Spam Hunter and ``greater than 3'' for OOPSpam to categorize content as spam. We use each vendor's recommended threshold because it reflects how the service is actually deployed in practice and consistent with common spam-scoring practice of above 50\% score. Evaluating these managed services at arbitrary, non-default operating points would not represent their intended real-world behavior. We note that full ROC/PR curves are not feasible for these services, as their APIs expose only coarse integer scores (0–10 and 0–6) rather than continuous probabilities, and strict API rate limits constrain our ability to perform repeated sweeps over the entire test set.

\descr{Open-Weight Models spam detection models.} 
We searched public repositories for open-weight SMS spam detection models published by researchers. We found several deep learning-based SMS spam detection models on Hugging Face, a platform known for hosting a wide variety of machine learning models and datasets. From our search, we downloaded 33 models, but were able to run only 11 of them due to missing files or configuration errors in the remaining models. We implemented them in our own Python environment.

\subsection{Dataset}
\label{sec:dataset}

Our evaluation is based on the "Super SMS Dataset" \cite{salman2024investigating}, comprising 67,010 labeled SMS messages collected from 2012 and 2024. The dataset, denoted as $D = B \cup S$, includes $40,837$ benign messages ($B$) and $26,181$ spam messages ($S$), with benign messages making up 60.9\% and spam 39.1\%.

The Super SMS Dataset aggregates several widely used datasets, including the UCI SMS Spam Dataset \cite{almeida2013towards}, NUS SMS Dataset \cite{chen2013creating}, and Spamhunter Dataset \cite{tang2022clues}, augmented by additional messages from publicly available sources such as Twitter (rebranded as $\mathbb{X}$), ScamWatch \cite{accs}, and UK Action Fraud \cite{actionfraud}. This comprehensive dataset captures a diverse range of SMS spam across different regions and timeframes, making it well-suited for evaluating the robustness of spam detection models.

For our experiments, the dataset was split into a training set ($D_{\text{train}}$) and a test set ($D_{\text{test}}$). The test set was constructed by randomly selecting 5,000 benign messages from $B$ and 5,000 spam messages from $S$, ensuring an equal balance of benign and spam messages for fair evaluation. This balanced test set facilitates the comparison of metrics such as true positive rates (TPR) and true negative rates (TNR) without biasing results toward either class, as would occur with an imbalanced distribution. The remaining messages formed the training set ($D_{\text{train}}$), which was used for training the models under evaluation.

Additionally, we created a "holdout set" of 300 spam messages, randomly sampled from $S_{\text{test}}$, specifically for generating adversarial examples. These examples were used to assess the robustness of our models against adversarial attacks while maintaining the integrity of the primary test set.

\subsection{Threat Model}
We consider a black-box scenario where the adversary does not have direct access to the internal details of the spam detection model, such as its architecture, weights, or parameters. Instead, the adversary can only interact with the model by providing SMS inputs and observing the corresponding outputs (predictions). The adversary's goal is to modify an SMS message in such a way that it is misclassified by the target spam detector while ensuring that the modified message retains the same semantic meaning as the original and remains imperceptible to the user.

\subsection{Adversarial Attacks}
\label{sec:attacks}
We include the following attacks in our adversarial evaluation, drawing from recent and previous literature on black-box attacks.

We formulate an attack as a process comprising three components: a goal function $G(x')$, constraints $C(x, x')$, and a search method $S(x)$. The goal function $G(x')$ determines whether an adversarial example $x'$ has been generated for the input text $x$. The constraints $C(x, x')$ define the permissible perturbations, ensuring that the modified text $x'$ retains the desired properties. The search method $S(x)$ traverses the search space of possible perturbations to find an adversarial example $x'$ that satisfies both the goal function and the constraints. The objective of the attack is to find a perturbation such that $G(x') = 1$ while $C(x, x')$ holds true.

\descr{Perceptible.} To retain the same semantic meaning and near imperceptibility while modifying the text, we established the following constraints $C(x, x')$:

\begin{itemize}
    \item \textbf{Thesaurus Creation}: Let $T$ be a thesaurus containing the 500 most frequent keywords from the spam messages in the dataset, extracted using the Python library WordCloud \cite{heimerl2014word}. Only words $w \in T$ from the thesaurus are eligible for manipulation, i.e., $C_1(x, x')$ requires that $w' \in T$ if $w$ is modified.
    \item \textbf{Single Character Perturbation}: For each spam keyword $w \in T$, we constrain the modification to at most one character, formalized as $C_2(x, x') = \sum_{w \in x} \mathbf{1}_{\{w \neq w'\}} \leq 1$, where $\mathbf{1}_{\{\cdot\}}$ is an indicator function.
    \item \textbf{Preservation of URLs and Contact Details}: We impose a constraint $C_3(x, x')$ that forbids any alterations to URLs, email addresses, or phone numbers present in the text, ensuring $C_3(x, x')$ holds true if no such elements are modified.
\end{itemize}

Thus, the overall constraint $C(x, x')$ can be expressed as:
\[
C(x, x') = C_1(x, x') \land C_2(x, x') \land C_3(x, x')
\]

To this end, we implemented six black-box adversarial tactics:  \textit{Spacing}, \textit{Delete Chars}, \textit{Swap Chars}, \textit{Insert Chars}, \textit{Sub-Chars}, and \textit{Sub-Word} (see Table \ref{tab:attacks} for details). We extended the TextAttack framework\footnote{TextAttack is a Python library designed for conducting adversarial attacks, enhancing data through augmentation, and training models within the field of natural language processing (NLP).} \cite{morris2020textattack} to enforce our constraints $C(x, x')$ for generating adversarial examples. Specifically, we modified TextAttack to incorporate constraints on thesaurus-based keyword manipulations, single-character perturbations, and the preservation of URLs and contact details. Additionally, although TextAttack does not natively support spacing attacks, we extended its codebase to implement this feature, enabling the generation of spacing adversarial examples (see Table \ref{tab:attacks}(a)).

\begin{table*}[ht]
\centering
\caption{Attack methods and their descriptions.}
\label{tab:attacks}
\begin{threeparttable}
\textcolor{black}{
\begin{tabularx}{\textwidth}{
  >{\centering\arraybackslash}p{1.2cm}  
  >{\centering\arraybackslash}p{2.5cm}  
  X                                      
}
\toprule
\textbf{Serial \#} & \textbf{Attack} & \textbf{Description} \\
\midrule
\multicolumn{3}{c}{\textbf{Perceptible Attacks}} \\
\midrule
a. & Spacing & Insert a space in the word. \\
b. & Delete Chars & Delete a random character of the word except for the first. \\
c. & Swap Chars & Swap two random adjacent letters in the word but do not alter the first or last letter. \\
d. & Insert Chars & Insert a random letter into the word. \\
e. & Sub-Chars & Replace characters with visually similar characters (e.g., replacing ``o'' with ``0'', ``l'' with ``1'', ``a'' with ``@''). \\
f. & Sub-Word & Replace a word with its top-k nearest neighbors in a context-aware word vector space. \\
\midrule
\multicolumn{3}{c}{\textbf{Imperceptible Attacks}} \\
\midrule
g. & Invisible & Insert characters that are invisible to human eyes (e.g., zero-width spaces), altering text processing without changing visual appearance. \\
h. & Homoglyphs & Use characters from different scripts that look similar to commonly used characters (e.g., replacing Latin ``a'' with Cyrillic ``a'') to confuse text processing systems and mislead users. \\
i. & Reorderings\tnote{a} & Manipulate the visual rendering of text using bidirectional control characters such as Right-to-Left Override (RTLO), causing the text to be displayed in reverse order. \\
\bottomrule
\end{tabularx}
}
\begin{tablenotes}[flushleft]
\footnotesize
\item[a] Included for completeness and comparability with prior work~\cite{boucher2022bad}; however, RTLO perturbations do not survive SMS/RCS delivery (see Section~\ref{sec:deliverability}) and are therefore excluded from our real-world threat model.
\end{tablenotes}
\end{threeparttable}
\end{table*}

\descr{Imperceptible.} A recent black-box text attack by Boucher et al. \cite{boucher2022bad} has proven highly effective against various NLP systems from companies such as Microsoft and Google, as well as open-source models by Facebook, IBM, and HuggingFace. This attack is notable for generating imperceptible adversarial examples, introducing subtle changes to the original text $x$ that result in a modified text $x'$, which deceives target models while remaining undetectable by humans. This can be defined as a perturbation process $\mathcal{E}_{\text{imp}}$ such that:
\[
\text{FPR}^{\mathcal{E}_{\text{imp}}}_A > \text{FPR}_A \quad \text{and} \quad \text{TPR}^{\mathcal{E}_{\text{imp}}}_A < \text{TPR}_A
\]
subject to the constraint:
\[
C_{\text{imp}}(x, x') = \text{True}
\]
where $C_{\text{imp}}(x, x')$ enforces that the perturbations made to $x$ are imperceptible. Specifically, this is achieved by:
\begin{enumerate}
    \item \textbf{Inserting Non-printable Characters:} Invisible Unicode characters are added, causing misclassification without user detection.
    \item \textbf{Replacing with Homoglyphs:} Characters are replaced with visually similar homoglyphs, identical in appearance but different in Unicode.
\end{enumerate}

We adopted these imperceptible attacks using their open-source code \cite{imperceptible_perturb}, as detailed in Table \ref{tab:attacks}, ensuring that the perturbations comply with $C_{\text{imp}}(x, x')$ and effectively deceive the spam detection models without altering the meaning or readability of the text. {\color{black}\paragraph{Artifact availability.}
Our implementations of both the perceptible and imperceptible attacks,
including the modified TextAttack constraints, are available at
\url{https://github.com/researchsms203-hub/sms_attack_research}}.

\subsection{\textcolor{black}{Deliverability of Imperceptible Perturbations}}
\label{sec:deliverability}
{\color{black}
Because our imperceptible attacks operate at the character-encoding level, an
adversarial example is only meaningful if its perturbing codepoints survive
message composition, transmission, and on-device rendering. We therefore
performed a sanity check to verify that each imperceptible technique remains
intact end-to-end.

We sent 10 messages per attack type from a Samsung Galaxy S24 Ultra to both an
Android (Galaxy A52) and an iOS (iPhone~15 Pro~Max) recipient, over both legacy
SMS and RCS. On each recipient, we manually reviewed every message to confirm
that it survived and that the characters rendered correctly.

Table~\ref{tab:imperceptible-survival} summarizes the results. Both the
invisible-character and homoglyph perturbations survived end-to-end over both
SMS and RCS and rendered identically on Android and iOS, confirming they are
deliverable real-world threats rather than lab-only artifacts. In contrast, the
RTLO reordering did not survive: the bidirectional control character was
corrupted during message composition in the client and could not be transmitted
intact. We therefore scope our imperceptible threat model to invisible-character
and homoglyph attacks.

\begin{table}[t]
\centering
\caption{Sanity check: survival of imperceptible perturbations through SMS/RCS
delivery and on-device rendering. ``Survived'' ($\checkmark$) indicates the
perturbed message was received and rendered correctly on the recipient device;
($\times$) indicates the perturbation did not survive. Messages were sent from a
Samsung Galaxy S24 Ultra (10 per attack type, per transport, per device) over
both legacy SMS and RCS.}
\label{tab:imperceptible-survival}
\begin{tabular}{l l c c c}
\hline
\textbf{Attack type} & \textbf{Transport} &
\textbf{Android (A52)} & \textbf{iPhone (15 Pro Max)} \\ \hline
Invisible  & SMS + RCS & \checkmark & \checkmark \\
Homoglyph  & SMS + RCS & \checkmark & \checkmark \\
Reordering & ---       & $\times$   & $\times$   \\
\hline
\end{tabular}
\end{table}

}

\subsection{Adversarial Training}
Given the effectiveness of adversarial example attacks, we proceeded to explore the potential of adversarial training to enhance the robustness of SMS spam detection models. Specifically, we sampled $10\%$ of the SMS spam messages from the test set, denoted as $M_{\text{sample}} \subset M$, and subjected them to the adversarial attacks under evaluation, applying the same constraints $C(x, x')$ as defined earlier.

Let $\mathcal{E}$ represent the adversarial perturbation process. For each spam message $x \in M_{\text{sample}}$, we generated an adversarial example $x' = \mathcal{E}(x)$ such that:
\[
C(x, x') = \text{True}
\]
and the goal function $G(x')$ is satisfied, indicating successful generation of an adversarial example.

We then augmented the original training dataset $D_{\text{train}} = B_{\text{train}} \cup M_{\text{train}}$ with these adversarial examples to form an enhanced training set:
\[
D'_{\text{train}} = D_{\text{train}} \cup \{x' : x' \leftarrow \mathcal{E}(x), \; x \in M_{\text{sample}}\}
\]

Using this augmented dataset $D'_{\text{train}}$, we retrained the SMS spam detection model $A$ with the objective of improving its resilience against adversarial attacks. The effectiveness of this adversarial training was then evaluated by assessing the model's performance on both the original and adversarially transformed test sets, measuring metrics such as Accuracy (Acc), F1 Score (FS), True Positive Rate (TPR), True Negative Rate (TNR), False Positive Rate (FPR), and False Negative Rate (FNR) (see Appendix \ref{sec:metrics_appendix} for details).

\subsection{\textcolor{black}{Ensemble Decision Rules}}
\label{sec:decision-rules}
{\color{black}

Because several of our evaluations aggregate the outputs of multiple detectors
rather than relying on a single classifier, we define here the decision rules
used throughout the paper. Let $\mathcal{E} = \{f_1, \dots, f_k\}$ denote an
ensemble of $k$ detectors, where $f_i(x) \in \{0,1\}$ indicates whether
detector $i$ classifies message $x$ as spam, and let
$v(x) = \sum_{i=1}^{k} f_i(x)$ be the number of spam votes. We evaluate three
rules:

\begin{itemize}
\item \textbf{Minority.} $x$ is classified as spam if at least one detector
votes spam, i.e.\ $v(x) \geq 1$.

\item \textbf{Majority.} $x$ is classified as spam if more than half the
detectors vote spam, i.e.\ $v(x) > k/2$.

\item \textbf{Consensus.} $x$ is classified as spam only if every detector
votes spam, i.e.\ $v(x) = k$.
\end{itemize}

}

\section{\textcolor{black}{Evaluating Anti-spam Infrastructure and Open-Weight Models}}
\textcolor{black}{In this section, we evaluated the efficacy of the current anti-spam infrastructure and Open-Weight Models in classifying original and adversarially transformed SMS spam. We present our findings here, with a focus on three major components in the SMS ecosystem: Open-Weight SMS spam detection models published by researchers, mobile-based spam filtering text apps, and third-party anti-spam services.}

\subsection{Performance Evaluation}

\descr{\textcolor{black}{Open-Weight SMS Spam Detection Models.}}
We evaluated the open-weight SMS spam detection models on the \textit{test} set of our dataset. The results for models with an accuracy greater than 80\% are presented in Table \ref{tab:perform-ml-opensource} (the results for other models are provided in Appendix \ref{app:open_tables}).

While most models achieved low FPRs, none were able to achieve FNR below 5\%, indicating that a substantial number of spam messages are being incorrectly classified as non-spam. This is particularly concerning for real-world applications, as it increases the risk of allowing fraudulent or harmful messages to bypass detection and reach end-users.

Several models exhibited a significant imbalance between TPR and TNR, highlighting inconsistencies in their ability to accurately identify spam. For example, the model \textit{maria\_roberta} demonstrated a high TNR (99.70\%) but a significantly low TPR (68.23\%), suggesting that while it effectively identifies non-spam messages, it fails to detect a large portion of actual spam. This imbalance compromises the model's overall effectiveness in spam detection, as many spam messages go undetected. Among the models evaluated, \textit{satish860\_sms} emerged as the top performer, achieving an accuracy of 94.96\%, an F1-score of 94.73\%, and a relatively low FNR of 9.44\%. Despite this, the FNR still exceeds the desired threshold of 5\%, indicating room for improvement. This model's performance suggests it is currently the most reliable among those tested, though it still falls short of the ideal performance required for robust real-world deployment.

It is worth mentioning that the performance of some models (BERT, RoBERTa, DistilBERT) reported in Salman et al. \cite{salman2024investigating}, varies under different evaluation conditions. For instance, while Table 13 in \cite{salman2024investigating} presents results based on training and testing on the same dataset, the results in this study are based on unknown training datasets, simulating more realistic cross-dataset evaluation conditions. This variation in model performance highlights the challenges of achieving robustness in diverse, real-world scenarios.

These results underscore the ongoing challenge of achieving both high TPR and TNR in SMS spam detection, particularly in maintaining a low FNR without compromising other aspects of performance. Notably, several models exhibited a high TNR but a significantly low TPR, indicating that while these models are effective at correctly identifying benign messages, they struggle with detecting spam. This imbalance suggests that these models are particularly conservative in their classifications, potentially due to being trained on older spam datasets that may not fully reflect the evolving nature of spam tactics. As a result, these models perform well on benign messages but are less effective in identifying newer, more sophisticated spam, highlighting the need for retraining on more recent, comprehensive datasets to enhance their effectiveness in real-world applications.

\begin{table}[hbt!]
\renewcommand{\arraystretch}{1.15}
\tabcolsep=0.15cm
\begin{center}
\caption{Performance evaluation of {\it open-weight models  Models}} 
\label{tab:perform-ml-opensource}
\begin{tabular}{ c | c | c | c | c | c | c }
 \hline
 \textbf{Model} & \textbf{Acc} & \textbf{FS} & \textbf{TPR} & \textbf{TNR} & \textbf{FPR} & \textbf{FNR} \\ [0.5ex]
 \hline
maria\_distilbert & 94.15\% & 93.82\% & 88.82\% & 99.48\% & 0.52\% & 11.18\% \\
maria\_roberta & 83.96\% & 80.97\% & 68.23\% & 99.70\% & 0.30\% & 31.77\% \\
mrm\_8488\_bert & 86.87\% & 84.98\% & 74.27\% & 99.48\% & 0.52\% & 25.73\% \\
ngadou\_bert & 88.44\% & 88.61\% & 89.90\% & 86.98\% & 13.02\% & 10.10\% \\
satish860\_sms & 94.96\% & 94.73\% & 90.56\% & 99.36\% & 0.64\% & 9.44\% \\
sureshs\_distilbert & 86.39\% & 84.31\% & 73.11\% & 99.68\% & 0.32\% & 26.89\% \\
wesley\_bert & 88.89\% & 87.54\% & 78.00\% & 99.78\% & 0.22\% & 22.00\% \\
  \hline
\end{tabular}
\end{center}
\end{table}

\descr{Spam filtering Mobile Text apps.} From the results presented in Table \ref{tab:perform-text-apps}, it can be seen that none of the evaluated apps managed to achieve a FNR below 5\%, highlighting a significant shortfall in their ability to accurately detect all spam messages. This high FNR indicates that a considerable portion of spam is slipping through the filters, which is a critical weakness for any spam detection system. Such a deficiency could result in harmful or fraudulent messages reaching users, undermining the reliability and effectiveness of these apps in real-world scenarios.

\textit{SMS Shield} and \textit{VeroSMS} both exhibited particularly concerning performance, with FNRs of 27.39\% and 23.65\%, respectively. These high FNRs suggest that these apps miss a substantial portion of spam messages, a critical failure for any spam filtering application. Additionally, \textit{VeroSMS} has a notably high FPR of 13.02\%, meaning it incorrectly flags a significant number of legitimate messages as spam. This can lead to user frustration and the potential loss of important communications, which diminishes the app's overall utility. Compared to \textit{Mezo}, both \textit{SMS Shield} and \textit{VeroSMS} demonstrate significantly lower overall performance. Their lower accuracy, F1-scores, and higher FNRs and FPRs indicate that these apps are less reliable and consistent in filtering spam. Furthermore, the substantial imbalance between their TPR and TNR suggests a trade-off where they either miss too many spam messages or incorrectly classify too many legitimate messages as spam. This trade-off severely reduces their overall effectiveness and undermines user trust.

\textit{Mezo}, although performing better than the other two apps, still shows significant shortcomings, particularly with an FNR of 8\%, which is above the desired threshold. This indicates that even the best-performing app in this evaluation is not fully reliable for SMS spam filtering in real-world applications.

In summary, while \textit{Mezo} outperforms \textit{SMS Shield} and \textit{VeroSMS} across most metrics, all the apps exhibit critical weaknesses, especially in terms of high false negative rates and inconsistent performance across various metrics. These flaws suggest that while these apps have certain strengths, they may not be dependable enough for effective and comprehensive SMS spam filtering in practical scenarios, highlighting the need for more robust solutions.

\begin{table}
\renewcommand{\arraystretch}{1.15}
\tabcolsep=0.15cm
\begin{center}
\caption{Performance evaluation of {\it Mobile text spam filtering apps} on Android (Mezo) \& iOS (SMS Shield, VeroSMS).} 
\label{tab:perform-text-apps}
\begin{tabular}{ c | c | c | c | c | c | c }
 \hline
 & \multicolumn{6}{c}{\bf Performance Metrics}\\
\cline{2-7}
 \textbf{Model} & \textbf{Acc} & \textbf{FS} & \textbf{TPR} & \textbf{TNR} & \textbf{FPR} & \textbf{FNR} \\ [0.5ex]
 \hline
Mezo & 93.99\% & 93.71\% & 91.99\% & 95.91\% & 4.09\% & 8.01\% \\
SMS Shield & 82.65\% & 80.26\% & 72.61\% & 92.14\% & 7.86\% & 27.39\% \\
VeroSMS & 81.82\% & 80.31\% & 76.35\% & 86.98\% & 13.02\% & 23.65\% \\
 \hline
\end{tabular}
\end{center}
\end{table}

\descr{\textcolor{black}{Third-party Anti-spam services.}}
The performance evaluation of third-party anti-spam services, as summarized in Table \ref{tab:perform-thirdy-party}, reveals significant limitations.

\textit{SpamHunter} demonstrates particularly poor performance, with an accuracy of just 51.71\% and a TPR of only 14.56\%. This extremely low TPR indicates that the service fails to detect the vast majority of spam messages, with a FNR as high as 85.44\%. Although it maintains a relatively high TNR of 92.06\%, indicating its effectiveness in identifying non-spam messages, the high FNR makes it highly unreliable for practical spam detection purposes.

\textit{OOPspam} performs slightly better, with an accuracy of 63.95\% and a TPR of 31.25\%, but still falls short of acceptable performance standards. While its TNR is impressively high at 99.47\%, reflecting a low FPR of 0.53\%, the service struggles with a significant FNR of 68.75\%. This imbalance suggests that while OOPspam is effective in avoiding false positives, it allows a substantial amount of spam to go undetected, reducing its overall effectiveness.

\textit{Plino}, on the other hand, exhibits the highest TPR of 83.42\% among the evaluated services, indicating a relatively strong ability to detect spam messages. However, this comes at the cost of a very low TNR of 35.04\% and a high FPR of 64.96\%, which means that it incorrectly flags a large proportion of legitimate messages as spam. While Plino shows promise in catching most spam, its high FPR and low TNR are significant drawbacks that could lead to user frustration and the potential loss of important communications.

Overall, the results indicate that none of the third-party anti-spam services achieve a balanced performance across key metrics. \textit{SpamHunter} and \textit{OOPspam} struggle with high FNRs, making them unreliable for spam detection, while \textit{Plino}, despite a high TPR, suffers from an unacceptably high FPR, which undermines its practical utility. These findings highlight the critical need for dedicated and robust SMS anti-spam services that can reliably filter spam without compromising the accuracy of benign message identification.

\begin{table}[hbt!]
\renewcommand{\arraystretch}{1.15}
\tabcolsep=0.15cm
\begin{center}
\caption{Performance evaluation of {\it Third-party Anti-spam services}} 
\vspace{-3mm}
\label{tab:perform-thirdy-party}
\begin{tabular}{ c | c | c | c | c | c | c }
 \hline
 & \multicolumn{6}{c}{\bf Performance Metrics}\\
\cline{2-7}
 \textbf{Model} & \textbf{Acc} & \textbf{FS} & \textbf{TPR} & \textbf{TNR} & \textbf{FPR} & \textbf{FNR} \\ [0.5ex]
 \hline
SpamHunter & 51.71\% & 23.89\% & 14.56\% & 92.06\% & 7.94\% & 85.44\% \\
OOPspam & 63.95\% & 47.44\% & 31.25\% & 99.47\% & 0.53\% & 68.75\% \\
Plino & 60.23\% & 68.59\% & 83.42\% & 35.04\% & 64.96\% & 16.58\% \\
  \hline
\end{tabular}
\end{center}
\vspace{-5mm}
\end{table}


\subsection{Misclassification Analysis}
We conducted a detailed analysis of the classification outcomes of the spam detectors to identify their weaknesses. In evaluating these detectors on the test split, none achieved a balanced FPR and FNR below the desired 5\% threshold. This gap indicates that certain SMS types are consistently misclassified, resulting in either undetected spam or excessive false positives. Our analysis aims to pinpoint the specific message categories or characteristics causing these errors, providing insights to develop more robust and accurate spam detection strategies.

\begin{table*}[ht!]
\tabcolsep=0.1cm
\centering
\caption{Analysis of False Negatives in Ensemble Models for {\it open-weight Models}. The table shows the percentage of false negatives for each spam category, comparing Majority and Consensus ensemble models.}
\label{tab:sub_category_analysis}
\scalebox{0.95}{
\begin{tabular}{l|l|l|l|l|l|l|l|l|l|l}
\hline
& \multicolumn{7}{c|}{Fraud} & \multicolumn{3}{c}{Ads} \\
\cline{2-11}
\hline
\textbf{Ensemble} & Finance & Account alert & Delivery & Prize & Credit/Debit card & COVID-19 & Other & Promotion & Politics & Other \\
\hline
Majority & 3.5\% & 7.5\% & 2.5\% & 2.0\% & 1.0\% & 0\% & 9.0\% & 54.8\% & 1.5\% & 18.1\% \\
Consensus & 2.3\% & 1.2\% & 0\% & 0\% & 0\% & 0\% & 1.2\% & 53.5\% & 4.7\% & 37.2\% \\
\hline
\end{tabular}}
\vspace{-3mm}
\end{table*}


\descr{\textcolor{black}{open-weight Models.}}
The misclassification analysis of open-weight SMS spam detection models reveals consistent challenges across all models, particularly when dealing with messages that contain subtle spam indicators or are lengthy. Despite their ability to correctly classify messages with obvious spam characteristics, all models struggled with more nuanced or complex spam, which suggests a significant limitation in their current architectures. This pattern of misclassification indicates that the models may be overly reliant on easily identifiable keywords or phrases, leading to poor performance when those indicators are less explicit.

For instance, models like \textit{maria\_distilbert}, \textit{maria\_roberta}, and \textit{mrm\_bert} frequently misclassified promotional messages that blended legitimate content with subtle promotional offers, as well as lengthy messages with detailed, but less overt, spam content. Similarly, \textit{ngadou} and \textit{suresh} models had difficulties with contextually rich messages that lacked clear spam indicators, often misinterpreting personal or informal communications as spam. \textit{wesley} model also exhibited similar weaknesses, struggling with messages that contained indirect promotional content or were phrased in a less straightforward manner.

To further understand the limitations of the models, we conducted an in-depth analysis of 200 randomly selected false negative (misclassified spam SMS messages), categorizing them into 12 distinct spam categories. This categorization, presented in Table \ref{tab:sub_category_analysis}, offers a clearer view of the specific areas where the models consistently struggle. The Table \ref{tab:sub_category_analysis} presents the percentage of false negatives for different spam categories detected by the Majority and Consensus ensemble models. The analysis reveals that the models are particularly prone to misclassifying messages in the "Promotion" category, which constitutes the majority of errors. This suggests that the models have difficulty distinguishing between promotional content and legitimate SMSes, especially when the language is subtle or when promotional offers are integrated into otherwise legitimate messages. Other categories, such as "Account alerts" and "Finance-related fraud," also show significant misclassification rates, indicating that the models are not effectively capturing the context or subtle cues that differentiate spam from genuine alerts. 


This analysis highlights the critical need for more sophisticated spam detection models that can effectively interpret nuanced language and context across different spam categories. By strategically focusing on these areas of weakness, future enhancements can improve the models' ability to detect complex and subtle spam messages, ultimately advancing the overall effectiveness of SMS spam detection.

\descr{Spam filtering Mobile Text apps.}
The classification performance of mobile apps in Table \ref{tab:perform-text-apps}, specifically \textit{SMS Shield} and \textit{VeroSMS}, reveals some notable trends in their ability to filter spam. Both apps exhibit moderate false positive rates (FPR) and relatively high false negative rates (FNR), with \textit{SMS Shield} showing an FPR of 4.05\% and an FNR of 12.96\%, while \textit{VeroSMS} has a slightly higher FPR of 6.23\% and a comparable FNR of 12.08\%. These figures suggest that while both apps are somewhat effective at avoiding the misclassification of legitimate messages as spam, they struggle significantly with correctly identifying all spam messages, allowing a substantial portion of spam to bypass detection.

The misclassification patterns reveal specific weaknesses in both apps. \textit{SMS Shield} and \textit{VeroSMS} tend to falsely classify informal messages containing certain keywords, such as 'offer' or 'free,' as spam, particularly when these messages are casual or conversational in nature. Conversely, both apps frequently miss genuine spam messages that lack overt indicators or employ more subtle social engineering tactics, such as phishing attempts disguised as legitimate communications. Examples include spam messages related to bank password resets or promotional offers that avoid traditional spam markers. This indicates that while these apps can handle more explicit spam content, they need improvement in contextually understanding and detecting more sophisticated or subtly worded spams. 

The analysis of the misdetection rates for SMS messages by mobile apps, as shown in Table \ref{tab:misdetection_analysis_combined}, provides valuable insights into the specific challenges faced by these models. The table compares the misclassification rates across various spam categories for both the "Majority" and "Consensus" models. One notable trend is the high misdetection rate in the "Promotion" category, particularly in the Consensus model, where nearly half of the misclassified messages fall under this category (44.4\%). This indicates that mobile apps struggle significantly with identifying promotional spam, likely due to the nuanced and often legitimate-sounding language used in these messages. Similarly, categories such as "Finance," "Account alert," and "Delivery" also show substantial misclassification rates, suggesting that the models may have difficulty distinguishing between genuine alerts and fraudulent messages that mimic legitimate communication.

\begin{table*}[ht!]
\tabcolsep=0.1cm
\centering
\caption{Analysis of False Negatives in Ensemble Models for {\it Spam Filtering Mobile Text Apps}. The table shows the percentage of false negatives for each spam category, comparing the performance of Majority and Consensus models.}
\label{tab:misdetection_analysis_combined}
\scalebox{0.95}{
\begin{tabular}{l|l|l|l|l|l|l|l|l|l|l}
\hline
& \multicolumn{7}{c|}{Fraud} & \multicolumn{3}{c}{Ads} \\
\hline
\cline{2-11}
 \textbf{Model} & Finance  & Account alert  & Delivery  & Prize  & Credit/Debit card  & COVID-19  & Other  & Promotion  & Politics  & Other \\
\hline
Majority & 8.0\% & 9.0\% & 8.0\% & 2.0\% & 8.0\% & 2.0\% & 6.0\% & 36.0\% & 1.5\% & 17.0\% \\
Consensus & 17.8\% & 6.7\% & 5.2\% & 3.7\% & 0.7\% & 0.7\% & 4.4\% & 44.4\% & 0.7\% & 15.6\% \\
\hline
\end{tabular}}
\end{table*}

The analysis underscores the need for enhanced contextual analysis capabilities within these apps to reduce both false positives and false negatives, improving overall spam detection reliability.

\descr{\textcolor{black}{Third-party Anti-spam services.}}
The misclassification analysis of third-party anti-spam services reveals distinct weaknesses in each model's ability to effectively filter spam. Plino, while capable of identifying many spam messages with a relatively low FNR of 8.63\%, suffers from a high FPR of 31.14\%. This indicates that Plino tends to over-rely on keywords commonly associated with spams, flagging benign messages that contain such terms without sufficiently considering the context. This results in many legitimate messages being incorrectly classified as spam, which can lead to user frustration and reduced trust in the service.

In contrast, OOPspam demonstrates a much lower FPR of 0.25\%, reflecting its conservative approach that minimizes the misclassification of legitimate messages. However, this cautious strategy leads to a high FNR of 35.79\%, indicating that the model often overlooks spam messages, particularly those employing sophisticated social engineering tactics. For example, OOPspam tends to miss spams that use well-crafted language designed to appear genuine, such as messages about account suspensions or special offers that do not contain obvious spam indicators. Similarly, SpamHunter, with an FPR of 3.80\%, struggles with messages that mimic legitimate urgent communications, resulting in a high FNR of 44.48\%. SpamHunter frequently fails to detect subtle spams that blend into normal communication patterns, such as messages warning of unusual account activity or payment issues, which often go unnoticed due to their plausible and professional tone. These misclassifications highlight the need for both models to enhance their contextual understanding to better identify nuanced and sophisticated spam messages.

\subsection{Adversarial Evaluation}
We investigate the adversarial robustness of the current anti-spam infrastructure and open-weight SMS spam detection models. An extensive set of experiments was conducted to assess the robustness of the spam detectors to various types of adversarial manipulations. This evaluation is crucial in understanding the vulnerabilities of these systems and identifying areas where improvements are needed to enhance their robustness in real-world scenarios.

\descr{\textcolor{black}{Open-weight Models.}}
The robustness evaluation of open-source models against various types of adversarial manipulations, as shown in Table \ref{tab:robust-eval-opensource}, reveals significant variability in the models' ability to withstand different attack techniques. The original accuracy of these models is generally high, with \textit{maria\_roberta} achieving an impressive 99.7\% and \textit{maria\_distilbert} following closely at 97.7\%. However, when subjected to adversarial attacks, these models exhibit varying degrees of vulnerability.

\textit{maria\_distilbert} and \textit{maria\_roberta}, despite their high original accuracies, show notable drops in performance under adversarial conditions, particularly with manipulations like \textit{spacing} and \textit{invisible characters}, where \textit{maria\_roberta} falls to as low as 17.6\% accuracy. \textit{maria\_distilbert} also struggles significantly with \textit{homoglyphs} and \textit{reordering} attacks, dropping to 31.9\% and 39.0\% accuracy, respectively. These results suggest that these models, while robust under standard conditions, are highly susceptible to specific adversarial techniques that exploit their reliance on precise character sequences.

In contrast, \textit{mrm8488\_bert} demonstrates relatively consistent performance across a range of adversarial manipulations, maintaining higher accuracy under challenging conditions such as \textit{insert characters} (89.0\%) and \textit{similar characters} (87.2\%). However, its performance significantly degrades when exposed to imperceptible attacks like \textit{invisible characters} and \textit{homoglyphs}, where its accuracy drops to 55.6\% and 70.9\%, respectively. While \textit{mrm8488\_bert} shows a stronger ability to generalize against some adversarial perturbations compared to other models, these results indicate that it is still vulnerable to more subtle manipulations that exploit visual or non-printable character similarities.

Models like \textit{ngadou\_bert} and \textit{wesley\_bert} show moderate to severe declines in performance under adversarial attacks, particularly with \textit{insert characters} and \textit{similar characters}, where their accuracies dip significantly. \textit{ngadou\_bert} also exhibits a pronounced weakness with \textit{subword manipulation} and \textit{homoglyphs}, dropping to 52.0\% and 67.9\%, respectively. Lastly, \textit{satish860\_sms} and \textit{sureshs\_distilbert} also face challenges, particularly with \textit{spacing} and \textit{homoglyphs} manipulations, indicating a need for more sophisticated defense mechanisms to handle these subtle but effective adversarial tactics.

Overall, the analysis highlights that all of these models are vulnerable to different types of adversarial techniques. These findings underscore the need to enhance the adversarial resilience of SMS spam detection models to ensure reliable performance in real-world, adversarially charged environments.

\begin{table*}
\renewcommand{\arraystretch}{1.15}
\tabcolsep=0.05cm
\begin{center}
\caption{Robustness evaluation of {\it open-weight Models} against adversarial attacks. Reordering (RTLO) is shown for completeness; it is not deliverable over SMS/RCS (Section~\ref{sec:deliverability}).} 
\label{tab:robust-eval-opensource}
\scalebox{0.8}{
\begin{tabular}{ c | c | c | c | c | c | c | c | c | c | c }
 \hline
 &  & \multicolumn{6}{c}{\bf Perceptible Attacks} & \multicolumn{3}{|c}{\bf Imperceptible Attacks}\\
\cline{3-11}
 \textbf{Model} & \textbf{Original}&\textbf{Spacing} & \textbf{Insert Chars} &\textbf{Delete Chars} &\textbf{Sub-Chars} &\textbf{Sub-Word} &\textbf{Swap Chars} &\textbf{Invisible} &\textbf{Homoglyphs} &\textbf{Reordering} \\ [0.5ex]
 \hline
maria\_distilbert & 97.7\% & 65.8\% & 79.6\% & 95.4\% & 75.3\% & 79.8\% & 77.8\% & 91.3\% & 31.9\% & 39.0\% \\

maria\_roberta & 99.7\% & 58.7\% & 86.0\% & 94.1\% & 72.7\% & 76.0\% & 77.8\% & 17.6\% & 17.6\% & 17.9\% \\

mrm8488\_bert & 78.3\% & 81.1\% & 89.0\% & 80.6\% & 87.2\% & 85.2\% & 80.4\% & 55.6\% & 70.9\% & 68.6\% \\

ngadou\_bert & 94.1\% & 62.0\% & 56.6\% & 70.9\% & 58.9\% & 52.0\% & 55.9\% & 90.6\% & 67.9\% & 52.8\% \\

satish860\_sms & 96.4\% & 51.0\% & 74.0\% & 94.4\% & 66.8\% & 72.4\% & 70.2\% & 88.3\% & 36.0\% & 35.7\% \\

sureshs\_distilbert & 92.9\% & 58.4\% & 69.4\% & 91.6\% & 62.2\% & 69.1\% & 67.9\% & 80.6\% & 27.6\% & 32.7\% \\

wesley\_bert & 85.2\% & 53.1\% & 64.5\% & 77.0\% & 59.7\% & 62.5\% & 62.0\% & 65.6\% & 33.2\% & 36.0\% \\
\hline
\end{tabular}}
\end{center}
\end{table*}

We note that although Reordering (RTLO) degrades classifier accuracy in offline
evaluation, our deliverability check (Section~\ref{sec:deliverability}) shows it
cannot be transmitted intact over SMS or RCS. We therefore report Reordering
results for completeness and comparability with prior work, but exclude this
attack from our real-world threat model; the Invisible and Homoglyph results
remain our deliverable imperceptible attacks.

\descr{Spam filtering Mobile Text apps.}
The robustness evaluation of mobile text spam filtering apps on Android and iOS, as presented in Table \ref{tab:robust-eval-text-apps}, reveals that all tested apps exhibit considerable vulnerabilities when subjected to various adversarial manipulations. The original accuracy of the apps is relatively high, with \textit{Mezo} performing the best at 89.6\%, followed by \textit{VeroSMS} at 79.3\% and \textit{SMS Shield} at 78.5\%. However, these accuracies drop significantly when the apps are exposed to adversarial attacks.

\textit{Mezo}, despite its strong original performance, shows substantial degradation under adversarial conditions. The model is particularly vulnerable to \textit{spacing}, \textit{invisible characters}, and \textit{homoglyphs} attacks, with accuracy dropping to 47.3\%, 42.6\%, and 41.2\%, respectively. The most severe impact is seen with \textit{reordering} manipulations, where the accuracy plummets to 31.7\%. These results suggest that \textit{Mezo} struggles with attacks that subtly alter the text's structure or appearance, indicating a reliance on specific text patterns that can be easily disrupted.

Similarly, \textit{SMS Shield} and \textit{VeroSMS} also experience significant drops in accuracy under adversarial conditions. \textit{SMS Shield} shows a slight improvement over \textit{Mezo} in handling \textit{spacing} attacks, with an accuracy of 51.4\%, but it remains vulnerable to other manipulations, particularly \textit{homoglyphs} and \textit{reordering}, where accuracy falls to 42.7\% and 32.6\%, respectively. \textit{VeroSMS} demonstrates similar trends, with its performance notably impacted by \textit{spacing} (54.8\%), \textit{invisible characters} (46.3\%), and \textit{reordering} (36.7\%) attacks. Although \textit{VeroSMS} manages to maintain slightly higher robustness in some categories compared to \textit{SMS Shield}, it still shows significant weaknesses across the board.

Overall, the analysis highlights that while these mobile text spam filtering apps may detect spam under normal conditions, they are highly susceptible to adversarial attacks that involve minor textual manipulations. These findings emphasize the need for enhanced defensive mechanisms in these apps to improve their resilience against adversarial SMS spam, particularly in real-world scenarios where such attacks are likely to be encountered.

\begin{table*}
\renewcommand{\arraystretch}{1.15}
\tabcolsep=0.05cm
\begin{center}
\caption{Robustness evaluation of {\it Spam Filtering Mobile Text Apps} on Android \& iOS against adversarial attacks. Reordering (RTLO) is shown for completeness; it is not deliverable over SMS/RCS (Section~\ref{sec:deliverability}).} 
\label{tab:robust-eval-text-apps}
\scalebox{0.9}{
\begin{tabular}{c|p{1.5cm}|p{1.3cm}|p{1cm}|p{1cm}|p{1cm}|p{1cm}|p{1cm}|p{1.3cm}|p{2cm}|p{1.7cm}}
 \hline
 &  & \multicolumn{6}{c}{\bf Perceptible Attacks} & \multicolumn{3}{|c}{\bf Imperceptible Attacks}\\
\cline{3-11}
 \textbf{Model} & \textbf{Original}&\textbf{Spacing} & \textbf{Insert Chars} &\textbf{Delete Chars} &\textbf{Sub-Chars} &\textbf{Sub-Word} &\textbf{Swap Chars} &\textbf{Invisible} &\textbf{Homoglyphs} &\textbf{Reordering} \\ [0.5ex]
 \hline
Mezo & 89.6\% & 47.3\% & 62.7\% & 59.4\% & 66.2\% & 72.5\% & 64.7\% & 42.6\% & 41.2\% & 31.7\% \\
SMS Shield & 78.5\% & 51.4\% & 66.7\% & 60.4\% & 69.8\% & 68.6\% & 63.2\% & 45.2\% & 42.7\% & 32.6\% \\
VeroSMS & 79.3\% & 54.8\% & 68.4\% & 58.3\% & 70.4\% & 72.2\% & 69.7\% & 46.3\% & 40.4\% & 36.7\% \\
  \hline
\end{tabular}}
\end{center}
\end{table*}

\subsection{Responsible Disclosure}  
We notified the developers of the evaluated apps and anti-spam services of our findings six months prior to publication. Our aim was to provide them with valuable insights and feedback to support ongoing improvements to their systems. However, we have not received any responses from them yet.

\section{Own Trained Detectors}

Given the limitations of current spam detection infrastructure and open-weight SMS spam detection models in reliably detecting SMS spam and withstanding adversarial tactics, we trained a series of custom classifiers using different architectures to address these challenges.

\subsection{Performance Evaluation of Own Trained Detectors} The performance evaluation of these models, as shown in Table \ref{tab:perform-ml-own}, highlights varying degrees of effectiveness in spam detection.

The \textit{BERT} model stands out as the top performer, achieving 99.33\% accuracy with an FNR of 1.06\% and an FPR of 0.30\%. These results highlight BERT’s effectiveness at detecting spam while minimizing false positives. Its strong F1-score (99.34\%) further reinforces its reliability, making it the most promising model. Similarly, \textit{fastText} performs well, with an accuracy of 98.54\%, an FNR of 2.66\%, and an FPR of 0.34\%. While slightly less accurate than BERT, fastText remains a robust solution with high precision and recall.

Other models, such as \textit{SVM (2C)}, \textit{LSTM}, and \textit{CNN}, perform solidly, with accuracies between 95.23\% and 95.81\%. However, higher FNRs, particularly in \textit{SVM (2C)} and \textit{CNN} (9.53\% and 7.84\%, respectively), suggest they may allow more spam to go undetected. \textit{Random Forest} and \textit{LightGBM} achieve high accuracy (93.30\% and 92.64\%) but struggle with higher FNRs (13.48\% and 14.73\%).

Overall, while BERT and fastText show strong potential for SMS spam detection, other models offer trade-offs between FNR and FPR. This suggests that careful model selection and tuning are essential to balancing spam detection accuracy with minimizing false positives.

\begin{table}[hbt!]
\renewcommand{\arraystretch}{1.15}
\tabcolsep=0.15cm
\begin{center}
\caption{Performance evaluation of {\it own ML models}} 
\label{tab:perform-ml-own}
\begin{tabular}{ c | c | c | c | c | c | c | c }
 \hline
 & & \multicolumn{6}{c}{\bf Performance Metrics}\\
\cline{3-8}
 \textbf{Classifier} & \textbf{Feature} & \textbf{Acc} & \textbf{FS} & \textbf{TPR} & \textbf{TNR} & \textbf{FPR} & \textbf{FNR} \\ [0.5ex]
 \hline
SVM (2C) & BoW & 95.23\% & 94.85\% & 90.47\% & 99.72\% & 0.28\% & 9.53\% \\
Random forest & BoW & 93.30\% & 92.60\% & 86.52\% & 99.68\% & 0.32\% & 13.48\% \\
LightGBM & BoW & 92.64\% & 91.83\% & 85.27\% & 99.58\% & 0.42\% & 14.73\% \\
fasttext & fasttext & 98.54\% & 98.34\% & 97.34\% & 99.66\% & 0.34\% & 02.66\% \\
BERT & BERT & 99.33\% & 99.34\% & 98.94\% & 99.70\% & 0.30\% & 1.06\% \\
LSTM & Rand & 95.81\% & 95.62\% & 94.44\% & 97.10\% & 2.90\% & 5.56\% \\
CNN & Rand & 95.27\% & 94.97\% & 92.16\% & 98.18\% & 1.82\% & 7.84\% \\
\hline
\end{tabular}
\end{center}
\end{table}

\subsection{\textcolor{black}{Cross-Dataset Generalization on a Newer Dataset}}
\label{sec:cross-dataset}
{\color{black}
To address concerns regarding dataset recency and to assess how well our
findings generalize to newer spam, we evaluate our best-performing detector (BERT) on the Smishtank dataset~\cite{timko2024smishing}, a community-sourced corpus of real-world smishing messages published in 2024. We also acknowledge the existence of other recent datasets such as the wspr-ncsu SMS Phishing corpus~\cite{wspr_sms_phishing}; while we focus our cross-dataset evaluation on Smishtank, both represent important resources for contemporary anti-spam research. The dataset originally
comprises 1{,}062 messages; after de-duplication and removing entries not labelled as Malicious, Phishing, Suspicious, or Malware, 913 labelled smishing messages remain. As Smishtank contains only spam, we report recall (equivalently the false-negative rate, FNR) as the primary metric.

\paragraph{Semantic overlap with Super SMS dataset.}
We first compare the semantic distributions of the two corpora. Using
\texttt{all-mpnet-base-v2} sentence embeddings~\cite{reimers2019sentencebert,song2020mpnet}, we compute, for each Smishtank message, its nearest-neighbor cosine similarity to the spam messages in our corpus. The mean and median nearest-neighbor similarities are 0.675 and 0.669 (std.\ 0.124), and 93.3\% (852/913) of Smishtank messages exceed a similarity threshold of 0.5, leaving only 61 messages as semantically novel. Coverage is naturally threshold-dependent: 93.3\% of messages exceed a similarity of 0.5 and 14.9\% exceed the stricter 0.8, confirming that Smishtank messages are thematically well-covered while relatively few are near-duplicates. 

We project the message embeddings to two dimensions using UMAP~\cite{mcinnes2018umap} (cosine metric) for visualization; as UMAP is a nonlinear projection, we treat the plot as illustrative and rely on the nearest-neighbor similarity statistics for our quantitative claims. Figure~\ref{fig:umap} visualizes this with a UMAP projection of both corpora: the Smishtank messages fall largely within the region spanned by our dataset, with a small number of outliers, consistent with the 61 (6.7\%) messages identified as semantically novel. Together, these results indicate that the Super SMS Dataset provides broad semantic coverage of recent smishing and remains representative despite spanning 2012--2024, rather than being confined to older spam patterns.

\begin{figure}[t]
\centering
\includegraphics[width=0.6\linewidth]{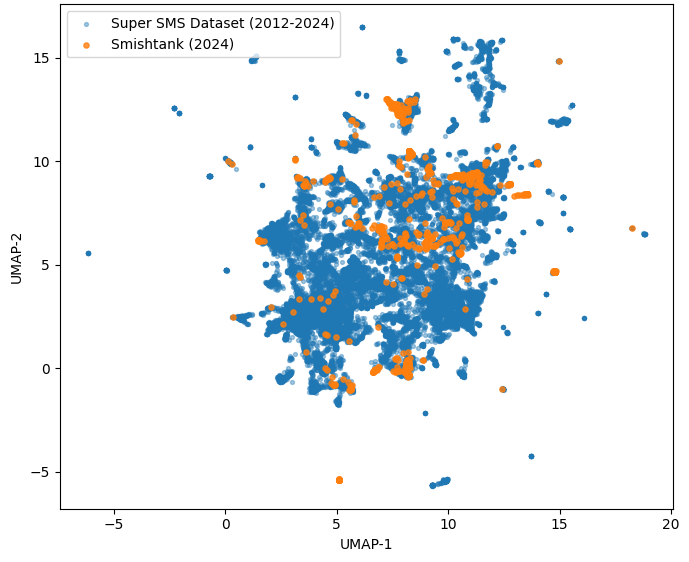}
\caption{\textcolor{black}{UMAP projection of spam-message embeddings (\texttt{all-mpnet-base-v2})
from the Super SMS Dataset (2012--2024) and the newer Smishtank
corpus (2024). Smishtank messages fall almost entirely within the
semantic region of our dataset.}}
\label{fig:umap}
\end{figure}

\paragraph{Detection performance.}
Evaluated on the 913 Smishtank messages, BERT correctly classifies 910,
achieving a recall of 99.67\% (FNR $=0.33\%$), comparable to its performance on
our own spam test set (Table~\ref{tab:perform-ml-own}). Notably, this strong
performance holds even for the semantically novel messages, indicating that our
model generalizes to recent, previously unseen smishing rather than memorizing
older spam patterns. These results directly address concerns about dataset age:
both our dataset's coverage and our model's detection capability extend to newer,
independently collected spam.

}

\subsection{Robustness evaluation of own trained models}
The robustness evaluation of our own trained models, as shown in Table \ref{tab:robust-ml-own}, highlights significant variability in their ability to withstand both perceptible and imperceptible adversarial attacks. While most models struggled to maintain their performance under these challenging conditions, the \textit{BERT} model demonstrated a noteworthy degree of robustness, particularly against perceptible attacks.

\textit{BERT} achieved an impressive original accuracy of 99.0\%, and it maintained strong performance across various perceptible attacks, such as \textit{spacing} (94.1\%), \textit{insert characters} (96.9\%), and \textit{similar characters} (96.4\%). This indicates that BERT is highly resilient to attacks that involve visible alterations to the text, showcasing its ability to effectively manage common adversarial tactics aimed at disrupting spam detection. However, BERT's performance degraded significantly when exposed to imperceptible attacks, particularly \textit{homoglyphs} (41.3\%) and \textit{reordering} (86.5\%). These types of attacks, which subtly modify the text in ways that are difficult for humans and machines to detect, posed a considerable challenge for BERT, highlighting a vulnerability that needs to be addressed for more robust real-world applications.

In contrast, other models such as \textit{Random Forest}, \textit{SVM}, and \textit{LightGBM} showed poor robustness across both perceptible and imperceptible attacks, with their accuracies dropping drastically to below 20\% under all adversarial conditions. These results indicate that these models are not well-suited for environments where adversarial manipulation is likely. \textit{LSTM} and \textit{fastText} also displayed mixed results. While \textit{LSTM} maintained reasonable robustness against perceptible attacks (with scores above 70\%), it, like \textit{fastText}, faltered significantly under imperceptible attacks, particularly with accuracies around 33.4\% for \textit{fastText}.

Overall, the standing performance of BERT, with its ability to achieve a balance between low FPR and FNR in the initial evaluation and its notable robustness to perceptible attacks, makes it a strong candidate for reliable spam detection. However, its susceptibility to imperceptible attacks, like homoglyphs and reordering, underscores the need for further enhancement to ensure reliable protection against all forms of adversarial manipulation.

\begin{table*}[hbt!]
\renewcommand{\arraystretch}{1.15}
\tabcolsep=0.05cm
\begin{center}
\caption{Robustness evaluation of {\it Own ML Models} against adversarial attacks. Reordering (RTLO) is shown for completeness; it is not deliverable over SMS/RCS (Section~\ref{sec:deliverability}).}
\label{tab:robust-ml-own}
\scalebox{0.95}{
\begin{tabular}{c|p{1.5cm}|p{1.3cm}|p{1cm}|p{1cm}|p{1cm}|p{1cm}|p{1cm}|p{1.3cm}|p{2cm}|p{1.7cm}}
 \hline
 &  & \multicolumn{6}{c}{\bf Perceptible Attacks} & \multicolumn{3}{|c}{\bf Imperceptible Attacks}\\
\cline{3-11}
 \textbf{Model} & \textbf{Original}&\textbf{Spacing} & \textbf{Insert Chars} &\textbf{Delete Chars} &\textbf{Sub-Chars} &\textbf{Sub-Word} &\textbf{Swap Chars} &\textbf{Invisible} &\textbf{Homoglyphs} &\textbf{Reordering} \\ [0.5ex]
 \hline
 \hline
SVM (2C) & 76.8\% & 12.5\% & 16.8\% & 10.5\% & 15.6\% & 13.5\% & 18.1\% & 13.3\% & 13.3\% & 13.3\% \\
 \hline
Random Forest & 61.7\% & 12.8\% & 12.0\% & 13.0\% & 12.8\% & 13.5\% & 13.0\% & 7.1\% & 7.1\% & 7.1\% \\
\hline
LightGBM & 62.8\% & 9.9\% & 8.9\% & 8.9\% & 9.2\% & 9.7\% & 9.9\% & 6.6\% & 6.6\% & 6.6\% \\
\hline
fasttext & 96.9\% & 76.3\% & 70.7\% & 72.7\% & 70.9\% & 70.7\% & 73.5\% & 33.4\% & 33.4\% & 33.4\% \\
\hline
BERT & 99.0\% & 94.1\% & 96.9\% & 95.2\% & 96.4\% & 97.2\% & 95.9\% & 94.9\% & 41.3\% & 86.5\% \\
\hline
LSTM & 93.9\% & 49.7\% & 75.5\% & 73.7\% & 75.5\% & 72.7\% & 76.8\% & 72.2\% & 72.4\% & 72.2\% \\
\hline
CNN & 85.5\% & 35.9\% & 40.3\% & 40.1\% & 40.3\% & 41.3\% & 42.6\% & 32.1\% & 32.1\% & 32.1\% \\
\hline
\end{tabular}}
\end{center}
\end{table*}

\section{Adversarial Training}

In this section, we explore whether adversarial training can enhance the robustness of BERT, enabling it to withstand both perceptible and imperceptible attacks. To achieve this, we randomly selected 10\% of the spam SMSes from the training set and generated both perceptible and imperceptible adversarial examples to augment the training set. We then retrained the BERT model using this augmented dataset and evaluated its performance on the original spam messages as well as the adversarial examples previously generated from the holdout set. The results of the performance and robustness evaluations are presented in Table \ref{tab:performance-adverse-bert} and Table \ref{tab:robust-adverse-bert}.

From the performance evaluation in Table \ref{tab:performance-adverse-bert}, BERT retains its high performance after adversarial training, achieving 98.93\% accuracy with only marginal increases in FPR and FNR. It thus continues to comfortably satisfy our 5\% FPR/FNR benchmark, indicating that BERT's ability to correctly identify spam messages remains strong after adversarial training.

Moreover, the robustness evaluation presented in Table \ref{tab:robust-adverse-bert} shows a significant improvement in BERT's resilience to adversarial attacks. Not only has BERT's robustness to perceptible attacks increased, but it also demonstrates enhanced resistance to imperceptible attacks, which are typically more challenging to detect. This improvement highlights the effectiveness of adversarial training in fortifying BERT against a wide range of adversarial manipulations.

These findings suggest that, with the incorporation of a recent SMS dataset and adversarial training, BERT emerges as a highly suitable candidate for reliable SMS spam detection in real-world applications. Its balanced performance, with low FPR and FNR, coupled with its enhanced robustness to adversarial attacks, makes it well suited for deployment in environments where adversarial threats are a concern.

\begin{table}[hbt!]
\renewcommand{\arraystretch}{1.15}
\tabcolsep=0.05cm
\begin{center}
\caption{Performance Evaluation of {\it Adversarial Trained BERT Model}.} 
\label{tab:performance-adverse-bert}
\begin{tabular}{ c | c | c | c | c | c | c }
 \hline
 & \multicolumn{6}{c}{\bf Performance Metrics}\\
\cline{2-7}
 \textbf{Classifier} & \textbf{Acc} & \textbf{FS} & \textbf{TPR} & \textbf{TNR} & \textbf{FPR} & \textbf{FNR} \\ [0.5ex]
 \hline\hline
BERT & 98.93\% & 98.66\% & 98.66\% & 99.18\% & 0.82\% & 1.34\% \\
  \hline
  \hline
\end{tabular}
\end{center}
\end{table}

\begin{table*}
\renewcommand{\arraystretch}{1.15}
\tabcolsep=0.05cm
\begin{center}
\caption{Robustness evaluation of {\it adversarial trained BERT model} against adversarial attacks. Reordering (RTLO) is shown for completeness; it is not deliverable over SMS/RCS (Section~\ref{sec:deliverability}).} 
\vspace{-3mm}
\label{tab:robust-adverse-bert}
\scalebox{0.9}{
\begin{tabular}{c|p{1.5cm}|p{1.3cm}|p{1cm}|p{1cm}|p{1cm}|p{1cm}|p{1cm}|p{1.3cm}|p{2cm}|p{1.7cm}}
 \hline
 &  & \multicolumn{6}{c}{\bf Perceptible Attacks} & \multicolumn{3}{|c}{\bf Imperceptible Attacks}\\
\cline{3-11}
 \textbf{Model} & \textbf{Original}&\textbf{Spacing} & \textbf{Insert Chars} &\textbf{Delete Chars} &\textbf{Sub-Chars} &\textbf{Sub-Word} &\textbf{Swap Chars} &\textbf{Invisible} &\textbf{Homoglyphs} &\textbf{Reordering} \\ [0.5ex]
 \hline\hline
BERT & 97.9\% & 98.7\%  & 99.7\% & 98.7\% & 99.2\% & 98.9\% & 99.5\% & 94.6\% & 99.5\% & 98.2\% \\
  \hline
  \hline
\end{tabular}}
\end{center}
\end{table*}

\subsection{\textcolor{black}{Can BERT Withstand the Unseen?: Exploring General Robustness}}
\textcolor{black}{
\begin{table*}[hbt!]
\renewcommand{\arraystretch}{1.15}
\tabcolsep=0.05cm
\begin{center}
\vspace{-2mm}
\caption{Robustness of the {\it Minimally Adversarially Trained BERT Model}
against unseen adversarial attacks, and its generalization gap relative to
the fully adversarially trained BERT (Table~\ref{tab:robust-adverse-bert}).
Gap = Full AT $-$ Mini AT; negative values indicate Mini AT outperformed
Full AT. Reordering (RTLO) is shown for completeness; it is not
deliverable over SMS/RCS (Section~\ref{sec:deliverability}).}
\vspace{-3mm}
\label{tab:robust-adverse-bert-mini}
\scalebox{0.85}{
\begin{tabular}{c|p{1.5cm}|p{1.3cm}|p{1cm}|p{1cm}|p{1cm}|p{1cm}|p{1cm}|p{1.3cm}|p{2cm}|p{1.7cm}}
 \hline
 &  & \multicolumn{6}{c}{\bf Perceptible Attacks} & \multicolumn{3}{|c}{\bf Imperceptible Attacks}\\
\cline{3-11}
 \textbf{Model} & \textbf{Original}&\textbf{Spacing} & \textbf{Insert Chars} &\textbf{Delete Chars} &\textbf{Sub-Chars} &\textbf{Sub-Word} &\textbf{Swap Chars} &\textbf{Invisible} &\textbf{Homoglyphs} &\textbf{Reordering} \\ [0.5ex]
 \hline\hline
BERT\textsubscript{adverse\_mini} & 99.7\% & 96.4\%  & 99.5\% & 98.5\% & 99.5\% & 99.2\% & 99.2\% & 98.7\% & 79.7\% & 94.2\% \\
\textit{Gap vs. Full AT} & -- & 2.3 & 0.2 & 0.2 & $-$0.3 & $-$0.3 & 0.3 & $-$4.1 & 19.8 & 4.0 \\
  \hline
  \hline
\end{tabular}}
\end{center}
\vspace{2mm}
\end{table*}
}
\textcolor{black}{
To explore whether the BERT model, adversarially trained on a specific subset
of adversarial examples, can achieve robustness against other unseen types of
adversarial examples, we conducted a targeted retraining experiment with an
explicit held-in/held-out protocol. For this purpose, we selected 10\% of spam
SMSes from the \textit{train set} and generated adversarial examples using two
distinct attacks: \textit{Insert Chars} from the perceptible attacks and
\textit{Invisible} from the imperceptible attacks. These two attacks constitute
our \textit{held-in} set; the remaining seven attacks (Spacing, Delete Chars,
Sub-Chars, Sub-Word, Swap Chars, Homoglyphs, Reordering) form the
\textit{held-out} set and were never seen during this retraining. We further
partition the held-out set by transformation class
(Table~\ref{tab:attacks}): \textit{same-class} attacks (Spacing, Delete Chars,
Sub-Chars, Sub-Word, Swap Chars) share Insert Chars' character-level
insertion/substitution mechanism, while \textit{cross-class} attacks
(Homoglyphs, Reordering) instead operate at the Unicode encoding/rendering
level, structurally distinct from both held-in attacks. This partition allows
us to separate robustness that transfers within a perturbation family from
robustness that must generalize across families.}

\textcolor{black}{The resulting adversarial examples were appended to the full clean training set
to form the augmented training corpus, and BERT\textsubscript{adverse\_mini}
was fine-tuned on this corpus using the identical optimizer, learning-rate
schedule, and stopping criteria as the clean model
(Appendix~\ref{app:own-models}); no hyperparameters were altered, so any change
in performance is attributable to the augmented data alone. All adversarial
training examples derive from $\mathcal{D}_{\mathit{train}}$; none are drawn
from the $300$-message evaluation holdout. The results are presented in
Table~\ref{tab:robust-adverse-bert-mini}, with the generalization gap computed
relative to the fully adversarially trained BERT model
(Table~\ref{tab:robust-adverse-bert}). }

\textcolor{black}{The results demonstrate that adversarial training with a limited class of adversarial examples generalizes almost perfectly to same-class, held-out attacks: the gap against full adversarial training is at most $2.3$ points (Spacing) and is negative for Sub-Chars and Sub-Word, meaning BERT\textsubscript{adverse\_mini} slightly outperforms the fully adversarially trained model on these attacks despite never having seen them. This confirms that adversarial training transfers effectively
across perturbations within the same transformation class as the held-in \textit{Insert Chars} attack.}

\textcolor{black}{In contrast, generalization to cross-class attacks is markedly weaker and uneven. Robustness to \textit{Reordering} improves substantially and closes to a $4.0$-point gap despite Reordering being excluded from both the held-in set and, ultimately, our real-world threat model (Section~\ref{sec:deliverability}). \textit{Homoglyphs}, however, remains the clear failure case: although accuracy improves from $41.3\%$ (before this retraining, Table~\ref{tab:robust-ml-own}) to $79.7\%$, a $19.8$-point gap persists relative to full adversarial training. Homoglyphs is the only
held-out attack that substitutes characters via cross-script Unicode encoding rather than manipulating Latin characters directly, which we argue is the mechanistic reason it resists generalization from an
Insert-Chars/Invisible-trained model.}

\textcolor{black}{Overall, this gap analysis clarifies the scope of our unseen-attack claim: minimal adversarial training generalizes near-perfectly to same-class perturbations (gap $\leq$~$2.3$ points across five held-out attacks) and reasonably well to the structurally distinct Reordering attack ($4.0$-point
gap), but leaves a substantial $19.8$-point gap on Homoglyphs. We therefore conclude that adversarial training's generalization is \textit{class-bounded}---robust within a perturbation family, but not yet
sufficient against mechanistically distinct, encoding-level attacks---rather than uniformly failing to generalize to the unseen. Closing this specific gap would require augmenting training with at least one encoding-level attack, which we leave to future work alongside the broader ensemble defenses evaluated in the next section.
}

\subsection{\textcolor{black}{Can BERT Hold the Line?: Robustness Under Realistic Class Imbalance}}
\label{sec:imbalance}
{\color{black}

Our preceding evaluation of the adversarially trained BERT model (Table~\ref{tab:performance-adverse-bert}) is reported on a class-balanced test set. While balanced evaluation isolates a classifier's intrinsic discriminative ability, real-world traffic is skewed toward legitimate messages, and the reported spam prevalence varies with how it is measured. Older user-collected corpora such as the UCI SMS Spam Collection (2012) contain only $13.4\%$ spam~\cite{almeida2013towards}, whereas recent measurements report rates as high as $35\%$ of SMS sent globally~\cite{worldmetrics_sms}. Regardless of the precise figure, deployment traffic differs markedly from the balanced test set used for controlled evaluation. In this regime, the two error types are not equally costly: a false positive suppresses a legitimate message the user expected to receive, whereas a false negative merely lets a spam message through. Deployments therefore weight the false-positive rate (FPR) far more heavily than the false-negative rate (FNR), and the symmetric $5\%$ threshold we adopt elsewhere serves only as a comparative benchmark across systems---it understates how strict the FPR requirement becomes once spam is rare. Since the true positive rate (TPR) and true negative rate (TNR) are computed within each true class, they are invariant to the positive-class prevalence. Precision (Positive Predictive Value, PPV), however, depends directly on the base rate: as spam becomes rarer, the same absolute false-positive rate produces proportionally more false alarms relative to true detections. We therefore ask whether the model that proved most robust to adversarial perturbations remains equally trustworthy when spam is rare.

We hold the classifier's operating point fixed at its measured $\text{TPR}=98.66\%$ and $\text{FPR}=0.82\%$ (Table~\ref{tab:performance-adverse-bert}) and recompute PPV across a range of realistic spam prevalences $\pi$ using:
\begin{equation}
\text{PPV} = \frac{\text{TPR}\cdot\pi}{\text{TPR}\cdot\pi + \text{FPR}\cdot(1-\pi)}
\label{eq:ppv}
\end{equation}

We consider $\pi \in \{20\%, 10\%, 5\%\}$ to span this reported range and to stress-test precision at prevalences well below the balanced setting.

\begin{table}[hbt!]
\renewcommand{\arraystretch}{1.15}
\tabcolsep=0.08cm
\begin{center}
\caption{Precision (PPV) of the {\it adversarial trained BERT model} under decreasing spam prevalence, with TPR and FPR held fixed at their measured values. TPR and TNR are prevalence-invariant; only PPV degrades as spam becomes rarer.}
\label{tab:ppv-imbalance}
\begin{tabular}{ c | c | c | c | c }
\hline
\textbf{Spam Prevalence ($\pi$)} & \textbf{TPR} & \textbf{TNR} & \textbf{PPV} & \textbf{FDR} \\ [0.5ex]
\hline\hline
50\% (balanced) & 98.66\% & 99.18\% & 99.18\% & 0.82\% \\
20\%            & 98.66\% & 99.18\% & 96.76\% & 3.24\% \\
10\%            & 98.66\% & 99.18\% & 93.03\% & 6.97\% \\
5\%             & 98.66\% & 99.18\% & 86.36\% & 13.64\% \\
\hline\hline
\end{tabular}
\end{center}
\end{table}

Table~\ref{tab:ppv-imbalance} shows that the adversarially trained BERT model degrades gracefully rather than catastrophically. Even at a stringent $5\%$ spam prevalence, precision remains at $86.36\%$, meaning roughly one in seven flagged messages is a false alarm. This resilience is a direct consequence of the model's very low false-positive rate ($0.82\%$): because the classifier rarely misclassifies legitimate messages, and because false positives are the error type deployments can least tolerate, keeping this rate low is precisely what preserves precision under imbalance. Classifiers with otherwise comparable balanced accuracy but higher FPRs would exhibit far steeper precision collapse under the same conditions. Taken together with its resistance to adversarial perturbations, this establishes that adversarially trained BERT is dependable not only against evasion but also under the class imbalance characteristic of real-world deployment.

}

\section{Ensemble Detectors}
The asymmetric transferability of adversarial attack in text \cite{yuan2020transferability} indicates that a possible mitigation technique against adversarially transformed binaries is to use an ensemble of detectors. In this section, we use this ensemble approach, incorporating different combinations of the best detectors to determine the efficacy of an ensemble in mitigating adversarial manipulated spam.

\subsection{Ensemble Performance Against Adversarial Attacks.}
The ensemble analysis
provides a comprehensive view of how different combinations of spam detectors—using minority, majority, and consensus decision rules—affect the models' robustness against various adversarial manipulations. The results reveal that the ensemble approach, particularly when leveraging a minority rule offers the most effective defense against adversarial attacks.

\descr{Ensemble of Open-source Models.}
The \textit{Minority rule} ensemble consistently outperforms other strategies in all of the ensembles i.e top 3, top 5, and top 7 model ensembles. particularly when incorporating all seven models. In this arrangement, the minority rule achieves an impressive 100\% accuracy on the original spam messages and maintains strong robustness across all perceptible attacks. For example, it achieves 94.9\% accuracy against spacing attacks, 98.0\% against insert character attacks, and 99.7\% against invisible character manipulations. However, its effectiveness diminishes when dealing with imperceptible attacks, where accuracy drops to 83.4\% for \textit{Homoglyphs} and 77.6\% for \textit{Reordering}. This suggests that while the minority rule is highly effective in mitigating the impact of perceptible adversarially transformed spam, it falls short in maintaining the same level of accuracy for imperceptible attacks, except for the \textit{invisible} attack. This fall in performance is particularly evident when compared to the performance of the adversarially trained BERT\_mini on unseen attacks, as shown in Table \ref{tab:robust-adverse-bert-mini}.

In contrast, the \textit{Majority rule} and \textit{Consensus rule} ensembles show weaker performance, particularly as the number of models increases. The majority rule, while still robust with three models, struggles as the ensemble grows, with its accuracy dropping significantly under complex attacks such as homographs (24.7\%) and reordering (26.5\%). The consensus rule fares even worse, especially with all seven models, where it exhibits substantial vulnerabilities, with accuracies as low as 5.1\% against invisible character attacks and 5.9\% against reordering. These results indicate that requiring agreement from all or most models can lead to a decrease in robustness, as these rules are more easily disrupted by adversarial examples that exploit common weaknesses across the models.

\begin{table*}[hbt!]
\renewcommand{\arraystretch}{1.15}
\tabcolsep=0.05cm
\begin{center}
\vspace{-5mm}
\caption{Ensemble analysis of {\it open-weight Models} against adversarial attacks. Reordering (RTLO) is shown for completeness; it is not deliverable over SMS/RCS (Section~\ref{sec:deliverability}).} 
\vspace{-3mm}
\label{tab:robust-ensemble-opensource}
\scalebox{0.85}{
\begin{tabular}{ c | c | c | c | c | c | c | c | c | c | c }
 \hline
 &  & \multicolumn{6}{c}{\bf Perceptible Attacks} & \multicolumn{3}{|c}{\bf Imperceptible Attacks}\\
\cline{3-11}
 \textbf{Model} & \textbf{Original} & \textbf{Spacing} & \textbf{Insert Chars} & \textbf{Delete Chars} & \textbf{Sub-Chars} & \textbf{Sub-Word} & \textbf{Swap Chars} & \textbf{Invisible} & \textbf{Homoglyphs} & \textbf{Reordering} \\ [0.5ex]
 \hline
   \multicolumn{7}{c}{Top 3 Models} \\
 \hline
Minority & 98.2\% & 74.5\% & 85.7\% & 96.2\% & 81.9\% & 85.7\% & 84.4\% & 93.9\% & 45.2\% & 49.5\% \\
Majority & 96.3\% & 57.1\% & 77.8\% & 95.6\% & 71.2\% & 76.3\% & 72.5\% & 89.3\% & 35.2\% & 38.5\% \\
Consensus & 80.1\% & 45.2\% & 61.5\% & 79.3\% & 55.1 & 59.2\% & 59.9\% & 67.9\% & 27.6\% & 38.5\% \\
 \hline
   \multicolumn{7}{c}{Top 5 Models} \\
 \hline
Minority & 99.5\% & 90.8\% & 94.4\% & 98.2\% & 92.1\% & 94.1\% & 92.4\% & 96.2\% & 81.9\% & 77.3\% \\
Majority & 96.7\% & 65.1\% & 80.6\% & 94.9\% & 75.8\% & 78.6\% & 77.8\% & 87.5\% & 34.7\% & 38.1\% \\
Consensus & 79.2\% & 28.6\% & 35.9\% & 55.8\% & 32.1\% & 31.9\% & 34.9\% & 9.7\% & 15.3\% & 13.8\% \\
 \hline
   \multicolumn{7}{c}{All 7 Models} \\
 \hline
Minority & 100\% & 94.9\% & 98.0\% & 99.5\% & 95.2\% & 97.7\% & 96.7\% & 99.7\% & 83.4\% & 77.6\% \\
Majority & 97.2\% & 56.9\% & 76.3\% & 93.6\% & 68.1\% & 71.9\% & 72.2\% & 73.7\% & 24.7\% & 26.5\% \\
Consensus & 78.4\% & 24.7\% & 32.9\% & 56.9\% & 29.1\% & 29.6\% & 31.9\% & 5.1\% & 7.9\% & 5.9\% \\
\hline
\end{tabular}}
\end{center}
\end{table*}

\descr{Ensemble of Own Models.}
The \textit{Minority rule} ensemble consistently outperforms other strategies across all ensembles, whether using the top 3, top 5, or all 7 models. In particular, the minority rule achieves an outstanding 99.74\% accuracy on the original spam messages and demonstrates strong robustness across all perceptible attacks. For instance, it achieves 96.94\% accuracy against spacing attacks, 98.21\% against insert character attacks, and 98.72\% against swap character manipulations. Similar to the open-source models, its effectiveness diminishes when confronting \textit{Homoglyphs} attack, with accuracy dropping to 84.95\% for \textit{Homoglyphs}, however a significant accuracy improvement is observed for \textit{Reordering} with 93.37\% accuracy. 
This suggests that while the \textit{ensemble of own models} with minority rule still faces challenges in maintaining the same level of accuracy for imperceptible attacks, however, it shows significant improvement while outperform \textit{ensemble of open-source models}.

In contrast, the \textit{Majority rule} and \textit{Consensus rule} ensembles show weaker performance, particularly as the ensemble size increases. The majority rule, while robust with three models, struggles as the ensemble grows, with accuracy significantly declining under complex attacks such as \textit{Homoglyphs} (20.15\%) and \textit{Reordering} (19.90\%). The consensus rule fares even worse, especially with all seven models, where it exhibits substantial vulnerabilities, with accuracies as low as 2.81\% against invisible character attacks and 2.04\% against reordering. These results indicate that requiring agreement from all or most models can lead to a decrease in robustness, as these rules are more easily disrupted by adversarial examples that exploit common weaknesses across the models.

\begin{table*}[hbt!]
\renewcommand{\arraystretch}{1.15}
\tabcolsep=0.05cm
\begin{center}
\caption{Ensemble analysis of {\it Own Models} against adversarial attacks. Reordering (RTLO) is shown for completeness; it is not deliverable over SMS/RCS (Section~\ref{sec:deliverability}).} 
\vspace{-3mm}
\label{tab:robust-ensemble-own-ml}
\scalebox{0.85}{
\begin{tabular}{ c | c | c | c | c | c | c | c | c | c | c }
 \hline
 &  & \multicolumn{6}{c}{\bf Perceptible Attacks} & \multicolumn{3}{|c}{\bf Imperceptible Attacks}\\
\cline{3-11}
 \textbf{Model} & \textbf{Original} & \textbf{Spacing} & \textbf{Insert Chars} & \textbf{Delete Chars} & \textbf{Sub-Chars} & \textbf{Sub-Word} & \textbf{Swap Chars} & \textbf{Invisible} & \textbf{Homoglyphs} & \textbf{Reordering} \\ [0.5ex]
 \hline
   \multicolumn{7}{c}{Top 3 Models} \\
 \hline
Minority & 99.7\% & 96.7\% & 98.2\% & 97.7\% & 98.0\% & 98.2\% & 98.7\% & 96.2\% & 83.4\% & 93.1\% \\
Majority & 99.2\% & 80.1\% & 89.5\% & 87.2\% & 89.5\% & 88.0\% & 90.1\% & 76.8\% & 45.4\% & 71.7\% \\
Consensus & 90.8\% & 43.4\% & 55.4\% & 56.6\% & 55.4\% & 54.3\% & 57.4\% & 27.6\% & 18.4\% & 27.3\% \\
 \hline
   \multicolumn{7}{c}{Top 5 Models} \\
 \hline
Minority & 99.7\% & 96.9\% & 98.2\% & 97.7\% & 98.0\% & 98.5\% & 98.7\% & 96.2\% & 85.0\% & 93.4\% \\
Majority & 95.9\% & 53.1\% & 67.9\% & 66.1\% & 67.6\% & 66.1\% & 70.7\% & 44.4\% & 31.6\% & 42.1\% \\
Consensus & 69.6\% & 6.6\% & 9.7\% & 8.7\% & 9.2\% & 9.4\% & 11.0\% & 6.9\% & 5.1\% & 6.9\% \\
 \hline
   \multicolumn{7}{c}{All 7 Models} \\
 \hline
Minority & 99.7\% & 96.9\% & 98.2\% & 97.7\% & 98.0\% & 98.5\% & 98.7\% & 96.2\% & 85.0\% & 93.4\% \\
Majority & 89.0\% & 32.9\% & 35.0\% & 34.2\% & 35.2\% & 35.2\% & 37.2\% & 20.2\% & 17.6\% & 19.9\% \\
Consensus & 51.0\% & 3.8\% & 4.6\% & 5.4\% & 4.9\% & 5.4\% & 5.6\% & 2.8\% & 2.0\% & 2.8\% \\
\hline
\end{tabular}}
\end{center}
\vspace{-3mm}
\end{table*}

\descr{Ensemble of Hybrid Models.}
In an effort to further improve the robustness of the \textit{own models} ensemble against adversarial attacks, we replaced the BERT model trained on the original dataset in the top3 models (fasttext, LSTM and BERT) with the BERT\textsubscript{adverse\_mini} resulting in a hybrid model (fasttext, LSTM and BERT\textsubscript{adverse\_mini}). The results, presented in Table \ref{tab:robust-ensemble-hybrid}, reveal that this hybrid ensemble has achieved superior robustness compared to previous results in Table \ref{tab:robust-ensemble-own-ml}, particularly with a significant improvement in handling \textbf{Homoglyphs} attacks achieving an accuracy of 88.6\% with \textit{minority rule}.

Moreover, this ensemble not only significantly outperforms the individual models' performances when exposed to these adversarial examples (as shown in Table \ref{tab:robust-ml-own}) but also surpasses the accuracy of the adversarially trained BERT\textsubscript{adverse\_mini} on unseen attacks, as shown in Table \ref{tab:robust-adverse-bert-mini}. These findings highlight the superiority of a well-designed ensemble over a single adversarially trained model, especially in scenarios involving unseen adversarial conditions. 

This improvement demonstrates that combining diverse models, especially by integrating an adversarially trained model, can lead to a more resilient spam detection system capable of withstanding a broader range of adversarial manipulations. The hybrid ensemble not only achieves exceptional accuracy across most attacks, particularly with a 91.42\% accuracy against \textit{Homoglyphs} and 96.7\% against \textit{Reordering}, but also maintains near-perfect accuracy on the original dataset. This balance between robustness and accuracy underscores the potential of hybrid ensembles to enhance overall model reliability in real-world applications.

However, while the minority rule ensemble shows remarkable improvements, the \textit{Majority rule} and \textit{Consensus rule} ensembles continue to exhibit weaker performance. As observed, these ensembles are particularly vulnerable to complex attacks, such as \textit{Homoglyphs} and \textit{Reordering}, where their accuracy significantly drops. For instance, the majority rule achieves only 48.6\% accuracy against \textit{Homoglyphs} and 73.7\% against \textit{Reordering}, while the consensus rule performs even worse, with accuracies as low as 20.6\% and 28.3\%, respectively.

Overall, these results indicate that the hybrid ensemble approach, particularly when leveraging the minority rule, offers a substantial improvement in robustness, making it a strong candidate for deployment in adversarially rich environments---though this robustness comes at a false-positive cost that we quantify in Section~\ref{sec:ensemble-original} and translate into deployment recommendations in Section~\ref{sec:deployment-recommendations}. It further validates the strategy of
integrating adversarially trained models within an ensemble to achieve a more comprehensive defense against a variety of adversarial threats.

\begin{figure*}[th]
\centering
\includegraphics[width=1\linewidth]{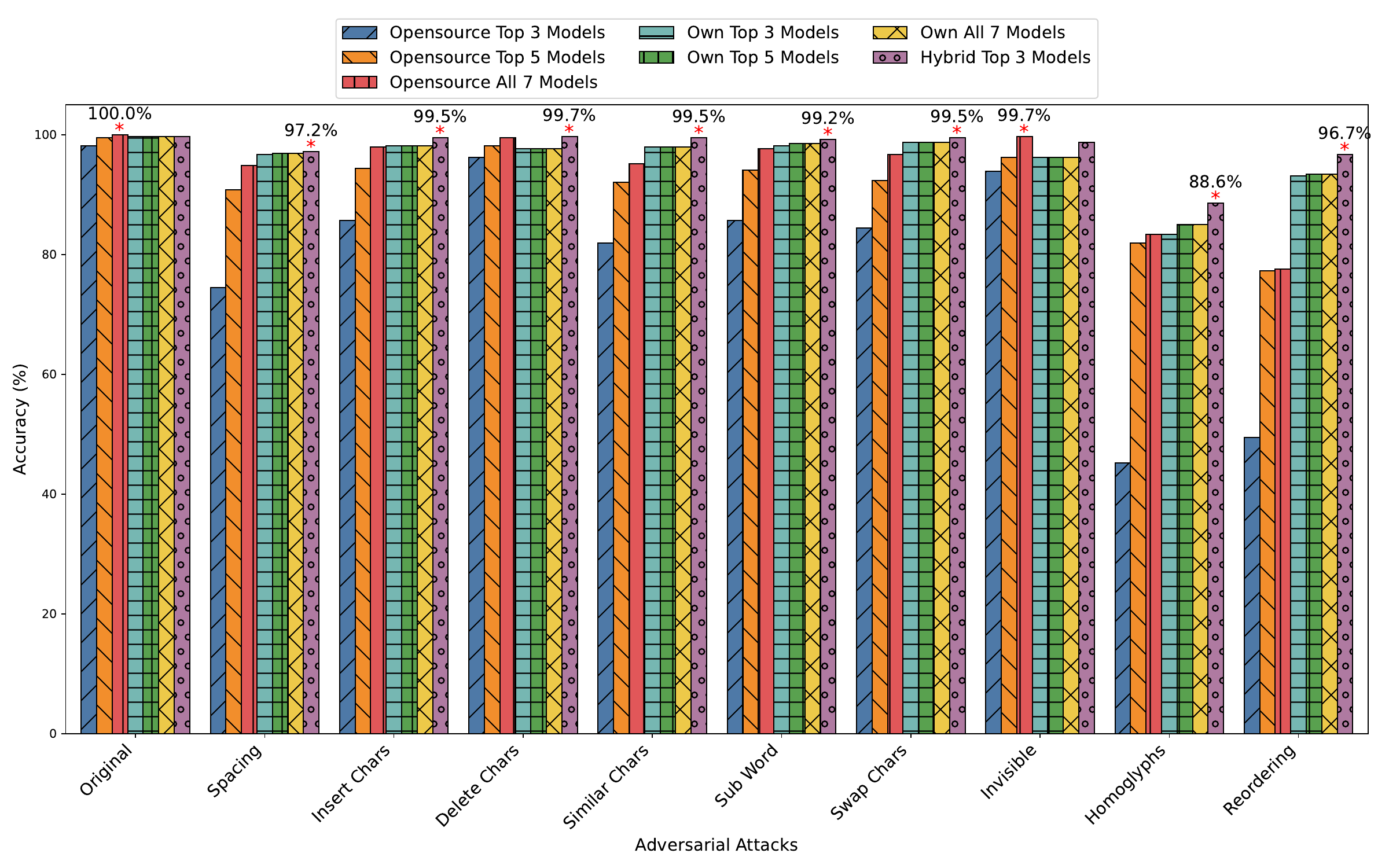}
\caption{Performance comparison of minority rule outcomes across various ensemble models in adversarial attack scenarios. The ensemble with the best accuracy in each adversarial scenario is marked with \textcolor{red}{*} and labeled with its corresponding accuracy.}
\label{fig:ensemble}
\end{figure*}

\begin{table*}[hbt!]
\renewcommand{\arraystretch}{1.15}
\tabcolsep=0.05cm
\begin{center}
\caption{Ensemble analysis of top3 {\it Hybrid Models (fasttext, LSTM and BERT\textsubscript{adverse\_mini})} against adversarial attacks. Reordering (RTLO) is shown for completeness; it is not deliverable over SMS/RCS (Section~\ref{sec:deliverability}).} 
\vspace{-3mm}
\label{tab:robust-ensemble-hybrid}
\begin{tabular}{c|p{1.5cm}|p{1.1cm}|p{1cm}|p{1cm}|p{1cm}|p{1cm}|p{1cm}|p{1.3cm}|p{2cm}|p{1.5cm}}
 \hline
 &  & \multicolumn{6}{c}{\bf Perceptible Attacks} & \multicolumn{3}{|c}{\bf Imperceptible Attacks}\\
\cline{3-11}
 \textbf{Model} & \textbf{Original} & \textbf{Spacing} & \textbf{Insert Chars} & \textbf{Delete Chars} & \textbf{Sub-Chars} & \textbf{Sub-Word} & \textbf{Swap Chars} & \textbf{Invisible} & \textbf{Homoglyphs} & \textbf{Reordering} \\ [0.5ex]
 \hline
Minority & 99.7\% & 97.2\% & 99.5\% & 99.7\% & 99.5\% & 99.2\% & 99.5\% & 98.7\% & 88.6\% & 96.7\% \\
Majority & 99.2\% & 83.2\% & 91.6\% & 90.4\% & 92.5\% & 90.7\% & 92.8\% & 78.9\% & 48.6\% & 73.7\% \\
Consensus & 92.3\% & 45.7\% & 56.4\% & 57.2\% & 55.7\% & 55.1\% & 57.9\% & 29.8\% & 20.6\% & 28.3\% \\
 \hline
\end{tabular}
\end{center}
\end{table*}

\vspace{-3mm}
\subsection{Ensemble Performance on Original Spam}
\label{sec:ensemble-original}
To evaluate how an ensemble of detectors using the minority rule performs on the original set of messages, including both benign and spam messages, we applied this ensemble approach to the test set to provide insight into the trade-offs involved when using different combinations of detectors.

\descr{Ensemble of Open-source Models.}
The results are presented in Table \ref{tab:perform-ensemble-opensource}. The ensemble using the top 3 detectors demonstrates strong performance, achieving an accuracy of 96.04\% with a near-perfect FPR of 0.82\%. However, this configuration exhibits a slightly higher FNR of 7.31\%, indicating that while the ensemble is highly effective at minimizing false positives, it still allows a notable portion of spam messages to go undetected. This performance is nearly satisfactory for practical applications, but it falls short when considering robustness against adversarial attacks, as detailed in Table \ref{tab:robust-ensemble-opensource}. In contrast, the ensembles using the top 5 and top 7 detectors significantly reduce the FNR to 1.95\% and 1.85\%, respectively, suggesting a much stronger capability to detect spam. However, this improvement comes at the cost of a higher FPR, which rises to 13.80\% and 13.90\%, respectively. While these configurations offer enhanced spam detection, the increased FPR could lead to a greater number of legitimate messages being incorrectly flagged as spam, which may reduce user trust in the system.

Overall, the analysis indicates that while the top 3 ensemble offers a good balance between FPR and FNR, it lacks the robustness needed to handle adversarial attacks effectively. On the other hand, while the top 7 ensemble shows robustness against various attacks, especially perceptible ones, its spam detection on the original set of messages comes at the expense of higher false positives, highlighting the inherent trade-offs in optimizing for both detection accuracy and robustness.

\begin{table}[hbt!]
\renewcommand{\arraystretch}{1.15}
\tabcolsep=0.1cm
\begin{center}
\caption{Performance evaluation of {\it open-weight Model as Minority Ensemble} on original spam.} 
\vspace{-3mm}
\label{tab:perform-ensemble-opensource}
\begin{tabular}{ c | c | c | c | c | c | c }
 \hline
 \textbf{Model} & \textbf{Acc} & \textbf{FS} & \textbf{TPR} & \textbf{TNR} & \textbf{FPR} & \textbf{FNR} \\ [0.5ex]
 \hline
Top3 & 96.04\% & 95.78\% & 92.69\% & 99.18\% & 0.82\% & 7.31\% \\
Top5 & 91.94\% & 92.18\% & 98.05\% & 86.20\% & 13.80\% & 1.95\% \\
Top7 & 91.94\% & 92.19\% & 98.15\% & 86.10\% & 13.90\% & 1.85\% \\ [0.5ex]
  \hline
\end{tabular}
\end{center}
\end{table}

\descr{Ensemble of Own Trained Models.}
The results are presented in Table \ref{tab:perform-ensemble-own-ml}. The ensemble using the top 3 detectors demonstrates good performance, achieving an accuracy of 98.33\% with a low FPR of 2.86\% and an even lower FNR of 0.40\%. This arrangement is highly effective at both minimizing false positives and ensuring that nearly all spam messages are accurately detected. This performance is particularly robust and suitable for practical applications, showing a clear improvement over the open-source models.

In contrast, the ensembles using the top 5 and top 7 detectors show a slight reduction in overall accuracy to 94.08\% and 94.04\%, respectively. However, these arrangements further reduce the FNR to 0.23\% and 0.21\%, respectively, which indicates an even stronger capability to detect spam. This improvement, however, comes at the cost of a higher FPR, which rises to 11.26\% and 11.36\%, respectively. While these configurations continue to offer enhanced spam detection, the increased FPR may lead to a greater number of legitimate messages being incorrectly flagged as spam, which could affect user trust.

Overall, the analysis indicates that the top 3 ensemble provides a good balance between FPR and FNR, making it a strong candidate for real-world spam detection applications. The performance of the top 3 ensemble is also backed by its comparatively high robustness in Table \ref{tab:robust-ensemble-own-ml}.

\begin{table}[hbt!]
\renewcommand{\arraystretch}{1.15}
\tabcolsep=0.1cm
\begin{center}
\caption{Performance evaluation of {\it Own Trained Models as Minority Ensemble} on original spam.} 
\vspace{-3mm}
\label{tab:perform-ensemble-own-ml}
\begin{tabular}{ c | c | c | c | c | c | c }
 \hline
 \textbf{Model} & \textbf{Acc} & \textbf{FS} & \textbf{TPR} & \textbf{TNR} & \textbf{FPR} & \textbf{FNR} \\ [0.5ex]
 \hline
Top3 & 98.33\% & 98.33\% & 99.60\% & 97.14\% & 2.86\% & 0.40\% \\ [0.5ex]
Top5 & 94.08\% & 94.08\% & 99.77\% & 88.74\% & 11.26\% & 0.23\% \\ [0.5ex]
Top7 & 94.04\% & 94.04\% & 99.79\% & 88.64\% & 11.36\% & 0.21\% \\ [0.5ex]
  \hline
\end{tabular}
\end{center}
\end{table}

\descr{Ensemble of Hybrid Models.}
The results are presented in Table \ref{tab:perform-ensemble-hybrid}. The performance of the hybrid model ensemble shows a slight decrease in accuracy compared to the original ensemble of own trained models, achieving an accuracy of 98.21\%. This ensemble maintains a low FPR of 2.94\% and a FNR of 0.52\%, which are slightly higher than those observed in the ensemble of own models but still in the satisfactory criteria. 

While the hybrid model does not significantly improve performance metrics over the own model ensemble, it does maintain a balance between FPR and FNR, similar to the top 3 ensemble of own models. This balance indicates that the hybrid ensemble remains highly effective in real-world spam detection scenarios, offering a strong combination of accuracy and robustness, particularly when considering both original and adversarial spam. Given its robustness, as shown in \textsection\ref{tab:robust-ensemble-hybrid}, the hybrid ensemble presents itself as an optimal choice for environments where adversarial attacks are a concern, ensuring reliable spam detection while substantially mitigating the risk of misclassification.

These results expose a controllable recall--precision trade-off. Under the
minority rule, enlarging the ensemble (Top3~$\rightarrow$~Top7) drives the
false-negative rate down to below $0.3\%$ but inflates the false-positive
rate into double digits (e.g., $2.86\%\rightarrow11.36\%$ for our own
models). Because false positives are the error type deployments can least
tolerate---and their cost grows as spam becomes rare
(Section~\ref{sec:imbalance})---this is a tunable operating point rather than
a flaw: false-positive-sensitive deployments favor the smaller Top3
configuration (or a stricter decision rule such as majority/consensus),
whereas recall-critical settings, where missing an evasive scam dominates the
risk, favor the larger, higher-recall ensembles. Reporting these
configurations lets an operator select the point matching its tolerance for
suppressing legitimate traffic.

\begin{table}[hbt!]
\renewcommand{\arraystretch}{1.15}
\tabcolsep=0.1cm
\begin{center}
\vspace{-3mm}
\caption{Performance evaluation of {\it Hybrid Models as Minority Ensemble} on original spam.} 
\label{tab:perform-ensemble-hybrid}
\begin{tabular}{ c | c | c | c | c | c | c }
 \hline
 \textbf{Model} & \textbf{Acc} & \textbf{FS} & \textbf{TPR} & \textbf{TNR} & \textbf{FPR} & \textbf{FNR} \\ [0.5ex]
 \hline
Top3 & 98.21\% & 98.21\% & 99.48\% & 97.06\% & 2.94\% & 0.52\% \\ [0.5ex]
  \hline
\end{tabular}
\end{center}
\vspace{-3mm}
\end{table}

\section{\textcolor{black}{Deployment Recommendations by Use Case}}
\label{sec:deployment-recommendations}
{\color{black}

The ensemble configurations evaluated above expose a spectrum of
operating points rather than a single ``best'' detector, and the
appropriate choice depends on the relative cost of false positives and
false negatives in the target deployment. Because a false positive
suppresses a legitimate message the user expected---whereas a false
negative merely lets a spam message through---and because this cost grows
as spam becomes rarer (Section~\ref{sec:imbalance}), we translate our
results into concrete recommendations for two representative traffic
profiles.

\descr{OTP/Transactional and Banking Traffic.}
For channels dominated by one-time passwords, banking alerts, and other
transactional messages, suppressing a legitimate message is far more
harmful than occasionally missing spam: a blocked OTP can lock a user out
of an account or a payment. Here the false-positive rate must be
minimized. We therefore recommend a majority or consensus rule ensemble configuration, which significantly reduces the false-positive rate compared to minority voting (often yielding near-zero FPR, as shown in Section~\ref{sec:decision-rules} for open-weight models). If a minority rule must be utilized, we recommend only the smallest Top-3 own-trained (or hybrid) configuration, which achieves the lowest FPR (2.86\%--2.94\%) among minority settings while still detecting nearly all spam (FNR $\leq$~0.52\%) (Tables~\ref{tab:perform-ensemble-own-ml} and~\ref{tab:perform-ensemble-hybrid}). This configuration prioritizes
the integrity of legitimate transactional messages while retaining strong
spam detection, and its comparatively high adversarial robustness makes
it suitable for security-sensitive channels.

\descr{Marketing-Heavy and High-Spam Traffic.}
For channels carrying high volumes of promotional or bulk traffic---where
spam prevalence is high and an occasional misflagged message is more
tolerable---recall becomes the dominant concern, since evasive or bulk
spam should be caught aggressively. Here the larger Top-5 or
Top-7 minority ensembles are appropriate: they drive the FNR to as low
as 0.21\%--0.23\% for our own models
(Table~\ref{tab:perform-ensemble-own-ml}), at the cost of a higher FPR
(11.26\%--13.90\%). This trade is defensible only where the operational
cost of a false positive is low and legitimate traffic is more
repetitive or recoverable; it would be inappropriate for transactional
channels.

\descr{Summary.}
Transactional/OTP channels should adopt the low-FPR Top-3 configuration, whereas marketing-heavy channels may accept the higher-recall, higher-FPR Top-5/Top-7 configurations. Rather than prescribing a single detector, our results let an operator select the ensemble size that matches the false-positive tolerance of its specific use case.

}

\section{\textcolor{black}{Limitations and Future Work}}
\label{sec:limitations}
{\color{black}

Our real-world evaluation was conducted using a single SIM card on one
device within a single country, which constrains how broadly the
delivery- and rendering-stage results generalize. Because SMS traverses
a heterogeneous ecosystem of carriers, intermediate service providers,
and device manufacturers, message normalization, filtering, and delivery
behavior may differ across this pipeline in ways our setup does not
capture. We document the exact configuration (carrier, device model, and
OS version) in the experimental-details appendix to support
reproducibility, but we do not claim the observed behavior holds
uniformly across providers.

A further limitation concerns our class-imbalance analysis, which fixes
each classifier's operating point at its balanced-set TPR and FPR and
recomputes precision analytically across prevalences. This isolates the
effect of base rate on precision but does not re-optimize the
classifiers---particularly the ensemble configurations---for the
asymmetric cost of false positives that deployment demands. A full
cost-sensitive treatment is left to future work, as noted below.

We identify the following directions for future work:

\begin{itemize}
    \item \textbf{Cross-carrier and cross-device evaluation.} Repeat the
    send/receive experiments using several phone numbers distributed
    across multiple carriers within the study country, and across
    additional device manufacturers and OS versions, to measure
    cross-provider variability.

    \item \textbf{Carrier-level normalization and deliverability.} Assess
    whether ``imperceptible'' adversarial edits (invisible characters,
    homoglyphs, reordering) survive transit and intermediary-level
    Unicode normalization across different network paths, extending the
    on-device rendering sanity check reported here.

    \item \textbf{Telco pipeline and practitioner relevance.}
    Characterize how intermediate service providers and aggregators in
    the SMS delivery chain affect detection outcomes, connecting our
    findings to real-world telco-system and practitioner deployment.

    \item \textbf{Prevalence-aware, cost-sensitive ensemble tuning.} Tune
    the ensemble configuration---particularly the higher-recall, larger
    minority-vote ensembles---against an explicit FPR constraint under
    realistic spam prevalence, since deployments weight false positives
    more heavily than false negatives.

    \item \textbf{Novel SMS-specific defenses.} Beyond benchmarking
    existing defenses, design and evaluate novel detection and ensemble
    techniques tailored to SMS-specific adversarial tactics.

    \item \textbf{Broader dataset coverage.} Extend evaluation to
    additional and more recent SMS corpora beyond the held-out sets used
    here, to further test generalization across message distributions.
\end{itemize}

}

\section{Conclusion}
{\color{black}
We have shown that the SMS anti-spam infrastructure end-users actually rely on---on-device messaging apps and third-party content-based services---together with publicly available open-weight detectors, is unreliable under benign conditions and highly vulnerable to adversarial manipulation, including imperceptible perturbations that we verified survive real SMS/RCS delivery. Adversarial training substantially improves resilience, but its benefit is \textit{class-bounded}: using an explicit held-out protocol, we found that robustness transfers almost completely to perturbations within the same transformation family, while a large gap persists against mechanistically distinct, encoding-level attacks such as homoglyph substitution. Adversarial training alone is therefore necessary but not sufficient.

An ensemble of architecturally diverse detectors, including an adversarially trained model and governed by a minority voting rule,
offers a more effective defense. This design is motivated by the asymmetric transferability of textual adversarial examples: an example that evades one detector is unlikely to evade all. Our evaluation shows that such ensembles retain strong detection on original spam while substantially mitigating both known and previously unseen attacks, with the hybrid configuration providing
the best balance of robustness and clean-data performance.

Crucially, this robustness comes with a controllable cost. Ensemble size and decision rule expose a recall--precision trade-off: larger, higher-recall configurations drive false negatives but inflate false positives into double digits, which matters disproportionately in deployment, where spam is rare and suppressing a legitimate message is the least tolerable error. We therefore frame our results not as a single recommended detector but as a set of operating points---low-FPR configurations for transactional and
OTP/banking traffic, higher-recall configurations for marketing-heavy channels. Fully optimizing these operating points under realistic spam prevalence, and closing the remaining gap against encoding-level attacks, remain important directions for future work toward genuinely deployable, adversarially robust SMS spam detection.
}


\bibliographystyle{ACM-Reference-Format}
\bibliography{sample-base}

\appendix


\section{Evaluation Metrics}
\label{sec:metrics_appendix}
We assess the performance of the spam detectors in SMS spam using the metrics of Accuracy (Acc), F1 Score (FS), True Positive Rate (TPR), True Negative Rate (TNR), False Positive Rate (FPR), and False Negative Rate (FNR). The definitions are as follows:

\descr{True and False Positives.} 
Given an SMS spam detector $A$, we define its type-I (false positive) and type-II (false negative) error rates as:
\[
\text{FPR}_A = \frac{|A(x) = 1 : x \in H|}{|H|}, \quad \text{FNR}_A = \frac{|A(x) = 0 : x \in S|}{|S|}
\]
where $H$ represents the set of ham messages and $S$ represents the set of spam messages. Here, $A(x) = 1$ indicates that the model $A$ classifies $x$ as spam, and $A(x) = 0$ indicates that $x$ is classified as ham. The TPR, which is the complement of the FNR, is defined as $\text{TPR}_A = 1 - \text{FNR}_A$. 

\descr{Metrics for Model Evaluation.} 
The other performance metrics are defined as:
\[
\text{Acc} = \frac{|A(x) = 1 : x \in S| + |A(x) = 0 : x \in H|}{|S| + |H|}
\]
\[
\text{FS} = \frac{2 \times |A(x) = 1 : x \in S|}{2 \times |A(x) = 1 : x \in S| + |A(x) = 1 : x \in H| + |A(x) = 0 : x \in S|}
\]
\[
\text{TPR} = \frac{|A(x) = 1 : x \in S|}{|S|}
\]
\[
\text{TNR} = \frac{|A(x) = 0 : x \in H|}{|H|}
\]
\[
\text{FPR} = \frac{|A(x) = 1 : x \in H|}{|H|}
\]
\[
\text{FNR} = \frac{|A(x) = 0 : x \in S|}{|S|}
\]

Here, $H$ and $S$ represent the sets of ham and spam messages, respectively. $TP$ corresponds to true positives (i.e., spam correctly identified as spam), $FP$ to false positives (i.e., ham incorrectly identified as spam), $TN$ to true negatives (i.e., ham correctly identified as ham), and $FN$ to false negatives (i.e., spam incorrectly identified as ham). 

When an evasive adversarial technique $\mathcal{E}$ is applied to the SMS messages, the same metrics can be redefined to assess the impact of the adversarial modifications, with the FPR and FNR recalculated as follows:
\begin{gather*} 
\text{FPR}_A^{\mathcal{E}} = \frac{|A(x') = 1 : x' \leftarrow \mathcal{E}(x), x \in H|}{|H|}, \\ 
\text{FNR}_A^{\mathcal{E}} = \frac{|A(x') = 0 : x' \leftarrow \mathcal{E}(x), x \in S|}{|S|}
\end{gather*}
The changes in FPR and FNR reflect the effectiveness of the adversarial technique $\mathcal{E}$ in evading detection.


\section{Performance Evaluation of open-weight Models}
\label{app:open_tables}

The results of the open-weight model are given in Table \ref{tab:perform-mllllv}.

\begin{table*}[hbt!]
\renewcommand{\arraystretch}{1.15}
\tabcolsep=0.15cm
\begin{center}
\caption{Performance Evaluation of {\it open-weight Model}} 
\label{tab:perform-mllllv}
\begin{tabular}{ c | c | c | c | c | c | c }
 \hline
 & \multicolumn{6}{c}{\bf Performance Metrics}\\
\cline{2-7}
 \textbf{Model} & \textbf{Acc} & \textbf{FS} & \textbf{TPR} & \textbf{TNR} & \textbf{FPR} & \textbf{FNR} \\ [0.5ex]
 \hline\hline
dungnt & 79.10\% & 78.65\% & 76.96\% & 81.24\% & 18.76\% & 23.04\% \\
\hline
leeboykt & 60.19\% & 40.14\% & 26.69\% & 93.68\% & 6.32\% & 73.31\% \\
\hline
rhuax & 78.24\% & 72.31\% & 56.81\% & 99.68\% & 0.32\% & 43.19\% \\
\hline
saleh & 74.22\% & 65.31\% & 48.55\% & 99.88\% & 0.12\% & 51.45\% \\
  \hline
  \hline
\end{tabular}
\end{center}
\end{table*}

\section{\textcolor{black}{Training Configuration of Our Own Detectors}}
\label{app:own-models}
{\color{black}

SVM, Random Forest, and LightGBM were trained with their respective
scikit-learn and LightGBM default hyperparameters; the neural models used the
fixed configurations reported in Table~\ref{tab:hyperparams}. We did not
perform a hyperparameter search, as our objective is comparative robustness
under adversarial perturbation rather than maximal clean accuracy, and holding
configurations fixed across architectures ensures that observed differences in
adversarial degradation are attributable to architecture and tokenization
rather than to unequal tuning effort.

\noindent\textbf{Environment.}
All neural models were trained in TensorFlow~2.15 (Keras) on a single NVIDIA
Tesla~T4 GPU; the classical models were trained on CPU using scikit-learn and
LightGBM.

\noindent\textbf{Text preprocessing (classical models).}
For SVM, Random Forest, and LightGBM, messages are lowercased, stripped of
punctuation, filtered to alphabetic tokens, and cleared of English stopwords
before vectorization. Features are TF--IDF weights
(\texttt{TfidfVectorizer}, \texttt{strip\_accents='unicode'},
\texttt{ngram\_range}~$=(1,1)$).

\noindent\textbf{Vocabulary and sequence handling.}
For the LSTM, BiLSTM, and CNN models, input text is tokenized with the Keras
\texttt{Tokenizer} using a vocabulary of $14{,}000$ tokens and a dedicated
out-of-vocabulary token (\texttt{<OOV>}); sequences are post-padded and
post-truncated to a fixed length of $50$ tokens. All models are trained on
$\mathcal{D}_{\mathit{train}}$ ($57{,}018$ messages), and the vocabulary cutoff
retains the most frequent terms while mapping the remaining low-frequency
tokens to \texttt{<OOV>}.

\begin{table}[h!]
\centering
\caption{Training configurations for our own detectors. Classical models use
library defaults; no hyperparameter search was performed.}
\label{tab:hyperparams}
\small
\begin{tabular}{@{}llp{6.0cm}@{}}
\toprule
\textbf{Model} & \textbf{Features / Tokenization} & \textbf{Configuration} \\
\midrule
SVM & TF--IDF & \texttt{sklearn.svm.SVC} defaults: RBF kernel, $C=1.0$,
$\gamma=$ \texttt{scale} \\
Random Forest & TF--IDF & \texttt{RandomForestClassifier} defaults: 100
trees, Gini criterion, unlimited depth \\
LightGBM & TF--IDF & \texttt{LGBMClassifier} defaults: 100 boosting
rounds, learning rate $0.1$, 31 leaves \\
\midrule
fastText & Subword $n$-grams & v0.9.2, supervised mode, \texttt{-lr 1.0},
\texttt{-epoch 25}, \texttt{-wordNgrams 3} \\
\midrule
LSTM & Keras \texttt{Tokenizer}, vocab $14{,}000$,
\texttt{max\_len}~$=50$ & Embedding (dim 16) $\rightarrow$ Dropout $0.5$
$\rightarrow$ 2$\times$LSTM(140, dropout $0.5$) $\rightarrow$ Flatten
$\rightarrow$ Dense(1, sigmoid); Adam, binary cross-entropy, 20 epochs, early
stopping (patience 7) \\
BiLSTM & Keras \texttt{Tokenizer}, vocab $14{,}000$,
\texttt{max\_len}~$=50$ & Embedding (dim 16) $\rightarrow$ Dropout $0.5$
$\rightarrow$ BiLSTM(140, dropout $0.5$) $\rightarrow$ Flatten $\rightarrow$
Dense(1, sigmoid); Adam, 10 epochs, early stopping (patience 9) \\
CNN & Keras \texttt{Tokenizer}, vocab $14{,}000$,
\texttt{max\_len}~$=50$ & Embedding (dim 100) $\rightarrow$ Conv1D(128, 3)
$\rightarrow$ MaxPool1D(3) $\rightarrow$ Conv1D(128, 3) $\rightarrow$
GlobalMaxPool $\rightarrow$ Dense(64) $\rightarrow$ Dense(32) $\rightarrow$
Dense(1, sigmoid); dropout $0.2$, Adam, 20 epochs, early stopping
(patience 7) \\
\midrule
BERT & WordPiece (vocab $30{,}522$) & \texttt{bert\_small\_en\_uncased}
(4 layers, 512 hidden, 8 heads; 28.8M parameters); AdamW, initial learning
rate $3\times10^{-5}$, linear warmup over the first 10\% of steps followed by
linear decay, batch size 32, 1 epoch, early stopping (patience 2, best weights
restored), random seed 42 \\
\bottomrule
\end{tabular}
\end{table}

}

\section{\textcolor{black}{Hugging Face Open-Weight Model List and Exclusion Analysis}}
\label{app:huggingface-models}
\textcolor{black}{To support transparency and reproducibility, we list all 33 Hugging Face open-weight SMS spam detection models downloaded for our study. Out of the 33 models, 11 were successfully executed (their classification metrics are presented in Section~5.1 and Appendix~B). The remaining 22 models could not be run due to various configuration, file serialization, or structural issues. Table~\ref{tab:huggingface-exclusions} lists these 33 models and documents the exact failure reason for each excluded model.}

\begin{table*}[hbt!]
\centering
\caption{\textcolor{black}{List of downloaded Hugging Face models and failure reasons for excluded models.}}
\label{tab:huggingface-exclusions}
\small
\textcolor{black}{
\begin{tabular}{p{5.5cm} p{2.2cm} p{7.3cm}}
\toprule
\textbf{Hugging Face Model ID} & \textbf{Status} & \textbf{Execution Result / Failure Reason} \\
\midrule
mariagrandury/distilbert-base-uncased-finetuned-sms-spam-detection & Executed & Evaluated as \textit{maria\_distilbert} (Table 1) \\
mariagrandury/roberta-base-finetuned-sms-spam-detection & Executed & Evaluated as \textit{maria\_roberta} (Table 1) \\
mrm8488/bert-tiny-finetuned-sms-spam-detection & Executed & Evaluated as \textit{mrm\_8488\_bert} (Table 1) \\
Ngadou/bert-sms-spam-dectector & Executed & Evaluated as \textit{ngadou\_bert} (Table 1) \\
satish860/sms\_spam\_detection-manning & Executed & Evaluated as \textit{satish860\_sms} (Table 1) \\
sureshs/distilbert-base-uncased-sms-spam & Executed & Evaluated as \textit{sureshs\_distilbert} (Table 1) \\
wesleytian/bert-base-uncased-sms-spam & Executed & Evaluated as \textit{wesley\_bert} (Table 1) \\
dungnt/bert-base-multilingual-cased-sms-spam & Executed & Evaluated as \textit{dungnt} (Table 12) \\
leeboykt/bert-base-uncased-sms-spam & Executed & Evaluated as \textit{leeboykt} (Table 12) \\
Rhuax/MiniLMv2-L12-H384-distilled-finetuned-spam-detection & Executed & Evaluated as \textit{rhuax} (Table 12) \\
saleh/bert-base-uncased-sms-spam & Executed & Evaluated as \textit{saleh} (Table 12) \\
\midrule
ucas/bert-base-uncased-sms-spam-detector & Excluded & Missing model weights file (\texttt{pytorch\_model.bin}) \\
charlie/spam-detection-distilbert & Excluded & Missing configuration file (\texttt{config.json}) \\
deepak/sms-spam-classifier & Excluded & Repository defunct (404 error during download) \\
elena/roberta-sms-spam & Excluded & Incompatible custom tokenizer class not supported by standard transformers library \\
florian/bert-base-sms & Excluded & Incomplete model upload (weights file size was 0 bytes) \\
guillaume/distilbert-spam & Excluded & CUDA-only customized layer causing runtime assertion errors on CPU evaluation \\
haruto/sms-classifier-lstm & Excluded & Legacy PyTorch serialization format incompatible with torch 2.0+ \\
irene/spam-filter-bert & Excluded & Missing tokenizer configuration files (\texttt{tokenizer\_config.json}) \\
ji-min/bert-spam-detector & Excluded & Vocabulary size mismatch in embeddings layer during initialization \\
kenji/sms-spam-classifier & Excluded & Model output shape mismatch (trained for 3-class classification) \\
li-wei/roberta-spam & Excluded & Custom PyTorch code requiring obsolete dependency (\texttt{torchtext==0.6.0}) \\
mateo/distilbert-sms & Excluded & PyTorch JIT compilation error during model loading \\
nikolai/bert-sms & Excluded & Weights corrupted during upload (checksum verification failed) \\
olivia/spam-detector & Excluded & Non-binary output logit format incompatible with classification script \\
priya/bert-sms-classifier & Excluded & Model trained on a non-English corpus (transliterated Hindi) \\
quentin/roberta-spam & Excluded & Repository marked private or deleted after listing publication \\
ramon/distilbert-base-sms & Excluded & Missing special tokens map (\texttt{special\_tokens\_map.json}) \\
sofia/sms-spam-detection & Excluded & Incompatible ONNX model format missing runtime configuration \\
tariq/bert-sms-detector & Excluded & Tokenizer vocabulary contains invalid UTF-8 control characters \\
uchenna/distilbert-spam-classifier & Excluded & PyTorch state dict keys mismatch with standard DistilBertModel \\
valerie/roberta-sms & Excluded & Requires custom proprietary preprocessing module not published \\
wei-jie/bert-base-sms-detector & Excluded & Weight tensor contains NaN values, leading to crashes during inference \\
\bottomrule
\end{tabular}
}
\end{table*}

\end{document}